\documentclass[11pt,draftcls,onecolumn,journal,table]{IEEEtran}
\usepackage{amsmath,amsfonts, amssymb}
\usepackage[many]{tcolorbox}
\usepackage{xcolor}
\usepackage{capt-of}
\usepackage{slashbox}
\usepackage{wrapfig}
\usepackage{algorithmic}
\usepackage{algorithm}
\usepackage{array}
\usepackage[caption=false,font=normalsize,labelfont=sf,textfont=sf]{subfig}
\usepackage{textcomp}
\usepackage{stfloats}
\usepackage{url}
\usepackage{verbatim}
\usepackage{graphicx}
\usepackage{cite}
\usepackage{hyperref}
\usepackage{booktabs}
\usepackage{bm}
\usepackage[framemethod=tikz]{mdframed}
\definecolor{bgblue}{RGB}{245,243,253}
\definecolor{ttblue}{RGB}{91,194,224}

\newtcolorbox{myboxii}[1][]{
  breakable,
  freelance,
  title=#1,
  colback=white,
  colbacktitle=white,
  coltitle=black,
  fonttitle=\bfseries,
  bottomrule=0pt,
  boxrule=0pt,
  colframe=white,
  overlay unbroken and first={
  \draw[red!75!black,line width=2pt]
    ([xshift=3pt]frame.north west) -- 
    (frame.north west) -- 
    (frame.south west);
  \draw[red!75!black,line width=2pt]
    ([xshift=-5pt]frame.north east) -- 
    (frame.north east) -- 
    (frame.south east);
  },
  overlay unbroken app={
  \draw[red!75!black,line width=2pt,line cap=rect]
    (frame.south west) -- 
    ([xshift=3pt]frame.south west);
  \draw[red!75!black,line width=2pt,line cap=rect]
    (frame.south east) -- 
    ([xshift=-5pt]frame.south east);
  },
  overlay middle and last={
  \draw[red!75!black,line width=2pt]
    (frame.north west) -- 
    (frame.south west);
  \draw[red!75!black,line width=2pt]
    (frame.north east) -- 
    (frame.south east);
  },
  overlay last app={
  \draw[red!75!black,line width=2pt,line cap=rect]
    (frame.south west) --
    ([xshift=3pt]frame.south west);
  \draw[red!75!black,line width=2pt,line cap=rect]
    (frame.south east) --
    ([xshift=-5pt]frame.south east);
  },
}

\newtheorem{remark}{Remark}

\def \bX {{\mathbf{X}}}
\def \bx {{\mathbf{x}}}

\def \bz {{\mathbf{z}}}

\def \bC {{\mathbf{C}}}
\def \bA {{\mathbf{A}}}

\def \cU {{\mathcal{U}}}
\def \bV {{\mathbf{V}}}

\def \bH {{\mathbf{H}}}

\def \br {{\mathbf{r}}}
\def \bv{{\mathbf{v}}}

\def \bp {{\mathbf{p}}}

\def \mR {{\mathbb{R}}}
\def \mE {{\mathbb{E}}}

\def \cH {{\mathcal{H}}}

\newtheorem{theorem}{Theorem}

\usepackage{multirow}

\usepackage[framemethod=tikz]{mdframed}
\begin{document}

\title{Disentangling Neurodegeneration with \\Brain Age Gap Prediction Models\\
{\LARGE {A Graph Signal Processing Perspective}}}

\author{Saurabh Sihag, Gonzalo Mateos, and Alejandro Ribeiro$^{\dagger}$,$^{\ast}$
\thanks{$^{\dagger}$Saurabh Sihag is with the Department of Electrical and Computer Engineering at the University at Albany, SUNY, Albany, NY (email: \href{ssihag@albany.edu}{ssihag@albany.edu}). Gonzalo Mateos is with the Department of Electrical and Computer Engineering at the
University of Rochester, Rochester, NY (email:\href{gmateosb@ece.rochester.edu}{gmateosb@ece.rochester.edu}). Alejandro Ribeiro is with the Department of Electrical and Systems Engineering at the University of Pennsylvania, Philadelphia, PA (email:\href{aribeiro@seas.upenn.edu}{aribeiro@seas.upenn.edu}).\\
$^{\ast}$ for the Alzheimer's Disease Neuroimaging Initiative (see Acknowledgement section).

}
}


\maketitle
Neurodegeneration is the progressive loss of structure or function of neurons in the brain. Reduction in cortical thickness or volume over time has been a workhorse metric to assess neurodegeneration in clinical settings\textcolor{black}{; see also Case Study 1 for a demonstration of cortical atrophy assessment in the context of Alzheimer's disease (AD) \textcolor{black}{relative to healthy individuals (HC group)}.} Naturally, visual inspection of T1-weighted brain magnetic resonance imaging (MRI) images and associated MRI quantification products are used along with other biological measurements to make a `subjective' assessment about the brain health of an individual. \textcolor{black}{These assessments tend to be subjective because they lack a deterministic relationship between an individual's health status and the absolute values of metrics observed within MRI scans~\cite{arrondo2022grey}.} Moreover, such methods cannot adequately account for the statistical complexities inherent within neuroimaging datasets that capture neurodegeneration. In particular, neurodegeneration is a characteristic of the healthy aging process and various neurological disorders~\cite{przedborski2003series}, exhibiting correlated patterns across brain regions. Such statistical factors motivate well the use of data-driven methods to characterize neurodegeneration.

Automating or improving the analyses of brain MRI images is appealing for several reasons: MRI is a non-invasive procedure and there is an untapped potential to reduce radiologists' missed detection error rates, leading to overall better patient treatment and outcomes, to name a few. In this article, we focus on the family of `Brain Age Gap Prediction' models. In simple terms, these models use machine learning (ML) algorithms to process neuroimaging data with the goal of predicting how much older the brain of an individual is relative to their chronological age; the difference being the so-termed brain age gap. Brain age gap prediction models have recently gained traction in digital health and personalized medicine, due to their promise of generating informative, yet compact, summary statistics of brain health for clinical use. Specifically, these models are hypothesized to leverage anomalous patterns associated with neurodegeneration in imaging data to yield a biomarker representative of the extent of neurodegeneration within an individual. This hypothesis has been corroborated in multiple recent studies, where the brain age gap (also, sometimes referred to as brain age delta) has been shown to be predictive of disease severity. \textcolor{black}{To offer the required background and context, }this tutorial begins with an overview of the brain age gap prediction algorithm and a survey of various existing studies that reinforce its relevance to characterizing neurodegeneration\textcolor{black}{; see `Brain Age Gap Prediction Models'.} 


\textcolor{black}{
Despite wide-ranging promising results, there are several challenges facing the practical deployment and generalizability of brain age gap prediction models in clinically meaningful settings (e.g., heterogeneous populations with distinct health conditions). Notable are methodological obscurities driven by the lack of a deterministic relationship (or conceptual justification) that ties the accuracy of the ML model for age prediction, to its usefulness in deriving a clinically significant brain age gap in neurodegeneration. In this context, we will review the relevant evidence from the 
literature 
and argue that such roadblocks to the practical adoption of brain age gap prediction stem from the opaqueness of the ML models used. 
The main contribution of this tutorial is to identify key mathematical principles that can help overcome the aforementioned challenges by bringing to bear graph signal processing (GSP).
}

Recent advances in GSP 
have introduced a breadth of \emph{principled} analytical tools for graph-structured (or correlated multivariate) data, \textcolor{black}{which match well with the intricacies of neuroimaging datasets. 
For instance, the cortical thickness features in Case Study 1 can be interpreted as \textit{graph signals} over some graph where nodes represent cortical brain regions (see Fig.~5). 
Edges are defined using the pairwise correlations 
between 
regional 
anatomical features
, i.e., the entries of the 
anatomical covariance matrix that will be affected by brain atrophy. 
We contend GSP offers a natural framework to study anatomical brain features and
to bridge key methodological gaps in brain age gap prediction.}
To this end, deep learning methods that build on GSP foundations take center stage. \textcolor{black}{In `GSP Foundations for Neuroimaging Data Analysis'} we review graph neural networks (GNNs) with convolutional layers and survey their theoretical properties, positioning them as an attractive tool for neuroimaging data analysis. We ground these discussions by introducing a GNN that is instantiated on an anatomical covariance matrix, called coVariance neural network (VNN). This way, a covariance matrix estimated from anatomical features derived from structural MRI is leveraged as a graph representation of signal structure.
\textcolor{black}{Crucially, we discuss the rich theoretical properties of VNNs and how they translate to unique operational capabilities that facilitate transparent and robust construction of brain age gap in neurodegeneration. Our discussions converge to a detailed overview of an explanation-driven brain age gap prediction pipeline; see `Towards Explainable Brain Age Gap Prediction from Structural MRI'. The exposition is supported by various case studies to demonstrate how key methodological obscurities in this application domain are overcome using VNN models. } 

All in all, this tutorial article elucidates the intellectual depth and clarifications added to brain age gap prediction algorithms via an interdisciplinary perspective rooted at the crossroads of GSP, ML theory, and network neuroscience. We conclude with brief discussions on adopting GSP for future studies in frontier applications such as domain-specific foundation models and neurodegenerative disease subtyping. 

\begin{mdframed}[hidealllines=true,backgroundcolor=gray!20]
{\bf Case Study 1: Cortical atrophy characterizes neurodegeneration in Alzheimer's disease (AD)}

\noindent \textcolor{black}{In this case study, we leverage the dataset from the ADNI study~\cite{wyman2013standardization} to demonstrate cortical atrophy in AD. This dataset consisted of three cohorts: (a) 206 healthy individuals (HC; age = $73.87 \pm 6.39$ years, $110$ females); (b) 372 individuals diagnosed with mild cognitive impairment (MCI; age = $72.26 \pm 7.61$ years, 160 females); and (c) 118 individuals diagnosed with AD (age =  $73.84 \pm 7.56$ years, 56 females). MCI diagnosis represents an early stage of loss of cognitive ability and is a precursor to AD. For each individual, $68$ cortical thickness features were available. These features are publicly available at \href{https://adni.loni.usc.edu/}{https://adni.loni.usc.edu/} and had been derived by processing T1-weighted structural MRI scans via Freesurfer software~\cite{fischl2012freesurfer}. The cortical thickness features were curated according to the Desikan-Killiany brain atlas~\cite{desikan2006automated}.}

\begin{minipage}{0.99\linewidth}
	\makeatletter
	\def\@captype{figure}
	\makeatother
	\centering
	\vspace{20pt}
	\includegraphics[width=0.8\linewidth]{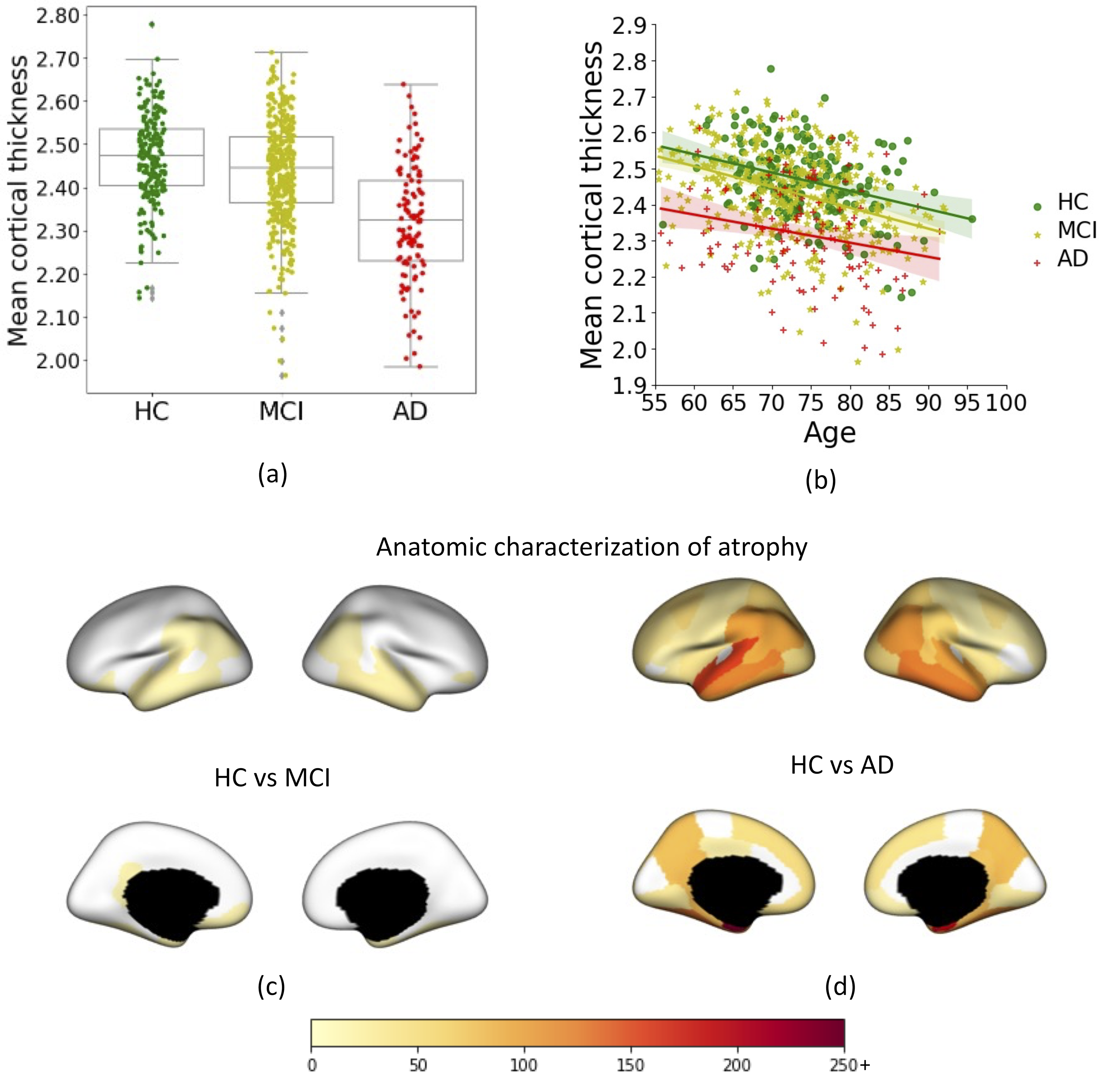}
	\caption{\footnotesize \textcolor{black}{(a) Distributions of mean cortical thickness across the 68 cortical regions in HC, MCI, and AD cohorts. (b) Scatterplot between the mean cortical thickness and age for the HC, MCI, and AD cohorts. The lines representing the linear fit for individual cohorts are also shown. (c) Brain atrophy as derived from cortical thickness in MCI cohort relative to HC cohort. (d) Brain atrophy as derived from cortical thickness in AD cohort relative to HC cohort. In (c) and (d), the $F$-values associated with statistically significant group differences in cortical thickness between MCI or AD groups and HC group as given by ANCOVA with age as covariate ($p$-value after Bonferroni correction $<0.05$) have been projected on the brain surface.} }\vspace{20pt}
\end{minipage}

\noindent \textcolor{black}{Figure~1 summarizes the characterization of neurodegeneration via brain atrophy as determined by cortical thickness features. Figure~1a illustrates the distributions of mean cortical metrics (across the whole brain) for the HC, MCI, and AD cohorts. With the reduction in mean cortical thickness representing }\textcolor{black}{ brain atrophy, it is apparent that the AD group exhibited higher brain atrophy than the HC group, with the MCI group falling in between them. Moreover, mean cortical thickness metrics for all groups exhibited negative linear relationships with age (Fig. 1b), suggesting that neurodegeneration was a characteristic of aging across all groups. Figures 1c and 1d provide the anatomic characterizations of brain atrophy in terms of cortical thickness features in MCI and AD groups. The MCI group exhibited statistically significant reduction in cortical thickness relative to HC group (ANCOVA with age as covariate, $p$-value after Bonferroni correction $<0.05$) in bilateral medial temporal lobe and temporo-parietal junction regions. Similar analysis revealed more prominent brain atrophy across a majority of brain regions in AD relative to HC in Fig. 1d, with the most prominent regions of atrophy including the bilateral entorhinal and medial temporal lobe. The contrast in the effect sizes of brain atrophy for MCI group in Fig. 1c and AD group in Fig. 1d is reasonable as the MCI diagnosis is typically a precursor of AD diagnosis. }
\end{mdframed}

\section*{Brain Age Gap Prediction Models}
Data from various neuroimaging modalities, including structural MRI, functional MRI, and positron emission tomography (PET), are able to capture the changes within various facets of the brain due to neurodegeneration and healthy aging~\cite{frisoni2010clinical}. We henceforth focus on structural MRI since it provides high-quality anatomical details of the brain and is among the most widely adopted modalities in clinical workflows. 
It also has potential diagnostic utility as features derived from structural MRI (brain atrophy, for example) can differentiate neurodegenerative conditions from healthy aging~\cite{koikkalainen2016differential}. The baseline for healthy aging is provided by the progressive anatomical and functional changes in the brain captured by neuroimaging datasets over the lifespan~\cite{lopez2013hallmarks}.

\begin{myboxii}
Within the landscape of data-driven ML algorithms that use structural MRI to identify neurodegeneration biomarkers, brain age gap prediction specifically targets the hypothesis that individuals can age \emph{biologically} at variable rates~\cite{peters2006ageing}.
\end{myboxii}

Neurodegeneration markers within structural MRI can be linked to the phenomenon of \textit{accelerated aging}, i.e., when the brain imaging data of an individual reflect patterns consistent with an advanced age relative to their current chronological age. For instance, brain atrophy is characterized by loss of cortical thickness and volume metrics, a characteristic of both healthy aging and neurodegeneration. Accelerated brain atrophy, often concentrated in specific brain regions, is a distinguishing feature of neurodegeneration relative to healthy aging. Hence, regions of the brain with accelerated atrophy relative to healthy individuals can be (statistically) perceived to have experienced accelerated aging. \textcolor{black}{This phenomenon is apparent in Fig. 1, where we observe a lower mean cortical thickness in the AD cohort relative to the HC group (Fig. 1a), with the reduction in cortical thickness concentrated in certain areas (Fig. 1d). In general, accelerated aging has been recognized as a predictor of morbidity and impairment~\cite{ferrucci2020measuring}. } 

\begin{figure}[t]
    \centering
    \includegraphics[width=\linewidth]{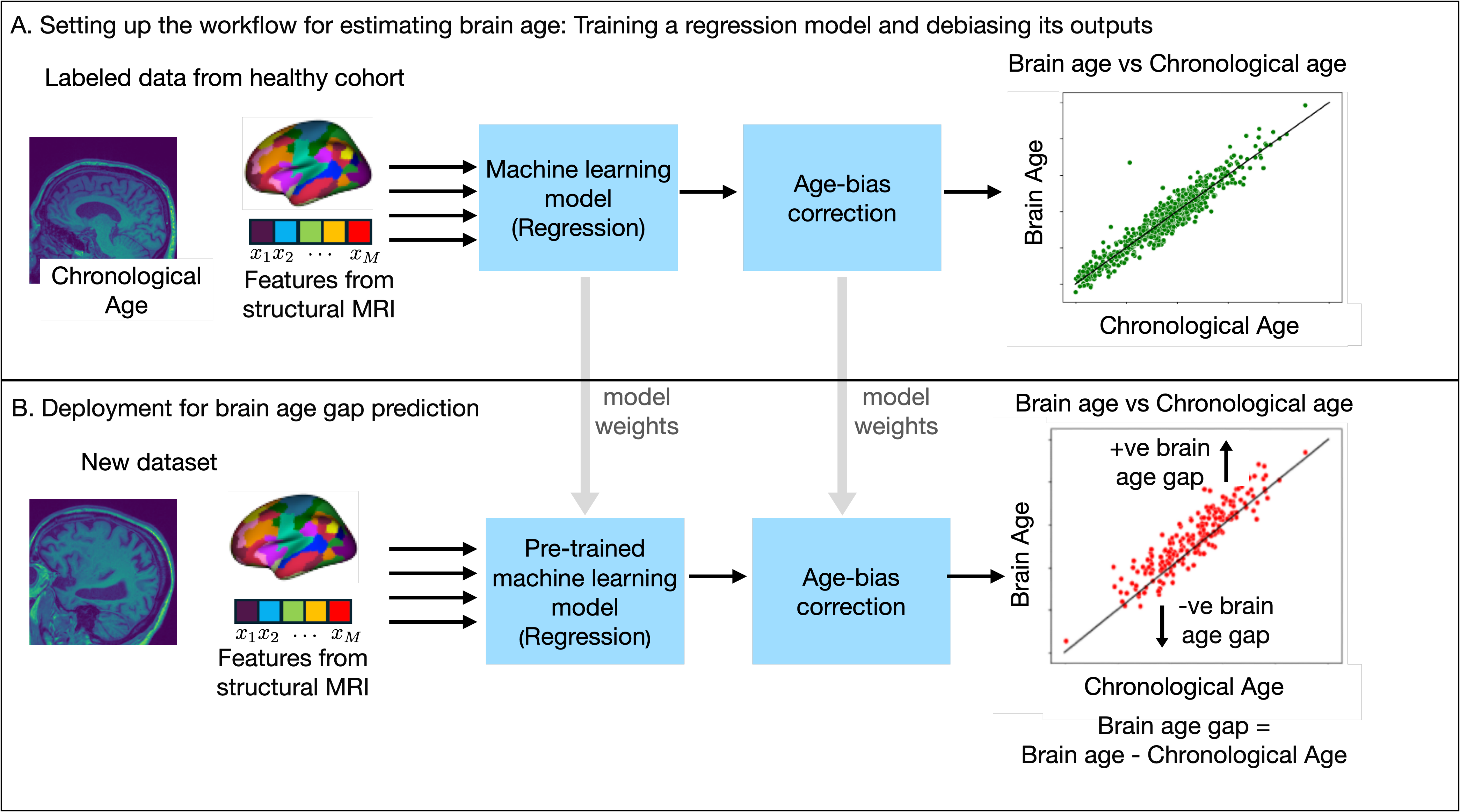}
    \caption{Schematic of brain age gap prediction using ML. (A) Neuroimaging data, formed by T1-weighted structural MRI scans, from a set of healthy individuals are labelled with their respective chronological age. Pre-processing pipelines using standard tools, such as Freesurfer, may be applied to extract relevant features from the MRI scan. A regression model is trained using the extracted features or the raw MRI scans, as preferred. The outputs of the ML model are then corrected for any age biases using an appropriate statistical correction procedure. Note that the age-bias correction is applied after training the regression model. The age-bias corrected outputs form the estimate of the brain age. (B) The trained ML model and its associated age-bias correction module can then be deployed to predict brain age using neuroimaging data pre-processed from a new dataset. The brain age gap is obtained as the difference between the predicted brain age and the chronological age.}
    \label{brain_age_calc}
\end{figure}

\textcolor{black}{Brain age gap prediction algorithms provide viable methods to quantify accelerated aging.} Specifically, they generate a compact scalar-valued representation of the biological age of an individual from features derived from structural MRI~\cite{baecker2021brain}. The metric of interest is the `Brain Age Gap', given by
\begin{align}
    \text{Brain Age Gap} \triangleq \text{Predicted Brain Age}  - \text{Chronological Age},
\end{align}
where predicted brain age is the estimate of biological age derived from neuroimaging data~\cite{COLE2017115}. In this context, brain age gap prediction is also often referred to as `brain age prediction' or `brain clock prediction', as the chronological age of an individual is usually known. Naturally, brain age gap prediction algorithms aim to extract traces of biological aging inherent in MRI scans. 


\begin{mdframed}[hidealllines=true,backgroundcolor=gray!20]
\textcolor{black}{{\bf How is brain age gap evaluated?}}

\noindent\textcolor{black}{Various facets of the schematic brain age gap prediction workflow in Fig. 2 are highlighted next. This workflow is primarily motivated by the hypothesis that an ML model \textcolor{black}{pre-trained to gauge} healthy aging can detect accelerated aging (i.e., infer brain age $>$ chronological age).}

\noindent\textcolor{black}{{\bf Data curation.} The training set for the ML model consists of the chronological age and the features derived from the structural MRI scans of a cohort of healthy individuals. The MRI scans may be pre-processed via image processing pipelines (such as Freesurfer~\cite{fischl2012freesurfer} and CAT12~\cite{farokhian2017comparing}) to extract meaningful features predictive of aging (for example, brain volume or thickness at each voxel). Moreover, the extracted features 
	may be organized anatomically according to a pre-selected brain atlas~\cite{desikan2006automated}. Some ML pipelines directly operated on the raw MRI scans~\cite{cole2017predicting}. } 

\noindent\textcolor{black}{{\bf Training \textcolor{black}{the ML model as} a regression model.} The features extracted from structural MRI are used as predictors in a regression model trained \textcolor{black}{to predict the chronological age of the healthy population}. 
	This pre-trained model provides an estimate $\hat y$ for an individual with chronological age $y$. The regression model is selected from the class of ML approaches suitable for multivariate data analyses, such as support vector regression, principal component analysis (PCA)-based regression, GNNs, or convolutional neural networks (CNNs). The loss function penalizes the (e.g., mean-squared error) deviation between the predicted outcome $\hat y$ and the chronological age $y$.}

\noindent\textcolor{black}{{\bf Age-bias correction.} \textcolor{black}{The predictions generated by the regression model for the healthy population are further evaluated for \emph{age bias}}, which may arise when the correlation between predicted age and chronological age is markedly smaller than $1$. In this scenario, the age of younger individuals may be overestimated while those of older individuals may be underestimated. This statistical bias correction could readily be applied with an appropriate linear model~\cite{de2020commentary}. A typical two-step age bias correction procedure to obtain the brain age prediction $\hat y_{\sf B}$ operates as follows~\cite{de2020commentary}:}

\noindent
{\bf Step 1.} Fit a linear model to the training set to estimate scalars $\omega$ and $\varrho$ in: $\hat y - y \sim \omega y + \varrho$.

\noindent
{\bf Step 2.} Evaluate brain age $\hat y_{\sf B}$ for an individual with chronological age $y$ and chronological age estimate $\hat y$ from their structural MRI features as follows:
\begin{align}\label{s2}
	\hat y_{\sf B} = \hat y - (\omega y + \varrho)\;. 
\end{align}
The difference between $\hat y_{\sf B}$ and $y$ is the brain age gap, henceforth denoted as  $\Delta$-Age, i.e.,
\begin{align}\label{ss}
	\Delta\text{-Age}\triangleq \hat y_{\sf B} - y.
\end{align}

\noindent\textcolor{black}{{\bf Deployment of the brain age gap prediction pipeline.} Features derived from the structural MRI scans of a new set of individuals (possibly with an unknown health status) can now be fed to the regression model with appropriate age-bias correction to generate the individual predictions of brain age. The difference between predicted brain age and chronological age quantifies the brain age gap. For example, if the predicted brain age of a 60-years-old individual is $70$ years, we say the brain age gap or $\Delta$-Age for this individual is $+10$ years.}	
\end{mdframed}

\subsection*{Brain age gap as a biomarker of neurodegeneration}

When trained on a healthy population, the ML model is expected to learn the statistical patterns of healthy aging. Using the learned representations of healthy aging as a reference point, the inferred $\Delta$-Age compactly captures accelerated aging if it has a smaller magnitude in a healthy population, and drifts toward a larger magnitude in the specific direction of $\Delta$-Age $>0$ for individuals with neurodegeneration~\cite{COLE2017115}. 

\begin{myboxii}
    The validity of brain age gap ($\Delta$-Age) as a biomarker of neurodegeneration hinges on its characterization of clinical markers of disease severity or risk and underlying biological processes.  
\end{myboxii}

The usefulness of $\Delta$-Age as a biomarker is often justified by demonstrating a combination of (i) higher $\Delta$-Age in neurodegeneration relative to the healthy population; and (ii) characterizing the clinical or biological markers of disease burden by post-hoc analyses of $\Delta$-Age; see Table~\ref{tab_review} for a summary of such representative works.  The former provides statistical evidence of accelerated aging in neurodegeneration, and (ii) is essential for interpretability. These aspects are illustrated in Case Study 2. 

\begin{mdframed}[hidealllines=true,backgroundcolor=gray!20]
\noindent  {\bf Case Study 2: Interpreting $\Delta$-Age as a biomarker in AD}	

\noindent
{\bf ML model.} \textcolor{black}{We consider GNN-based regression for brain age gap prediction from cortical thickness features in the ADNI dataset; see Case Study 1 and~\cite{sihag2024brain}. A GNN is chosen to exploit the correlation-induced information structure among the cortical-thickness features.} The model was trained to predict the chronological age of the cognitively healthy population from the OASIS-3 dataset~\cite{lamontagne2019oasis} (611 individuals, age = $68.38 \pm 7.62$ years, 351 females), using their cortical thickness features curated according to the Desikan-Killiany atlas.

\noindent{\bf $\Delta$-Age in AD.} The pre-trained model was used to predict $\Delta$-Age for different cohorts in the ADNI dataset described in Case Study 1~\cite{wyman2013standardization}. We also investigated the relationship between $\Delta$-Age and Clinical dementia rating- sum of boxes (CDRSB) metrics to assess the clinical interpretation of $\Delta$-Age. CDRSB is a clinical marker to stage dementia severity~\cite{o2008staging}. CDRSB for MCI (available for $200$ individuals, mean = 1.33, standard deviation  = 0.94) was substantially smaller relative to that for the AD cohort (available for $70$ individuals, mean = 5.61, standard deviation = 2.35). 

\begin{minipage}{0.99\linewidth}
	\makeatletter
	\def\@captype{figure}
	\makeatother
	\centering
		\vspace{20pt}
	\includegraphics[width=0.85\linewidth]{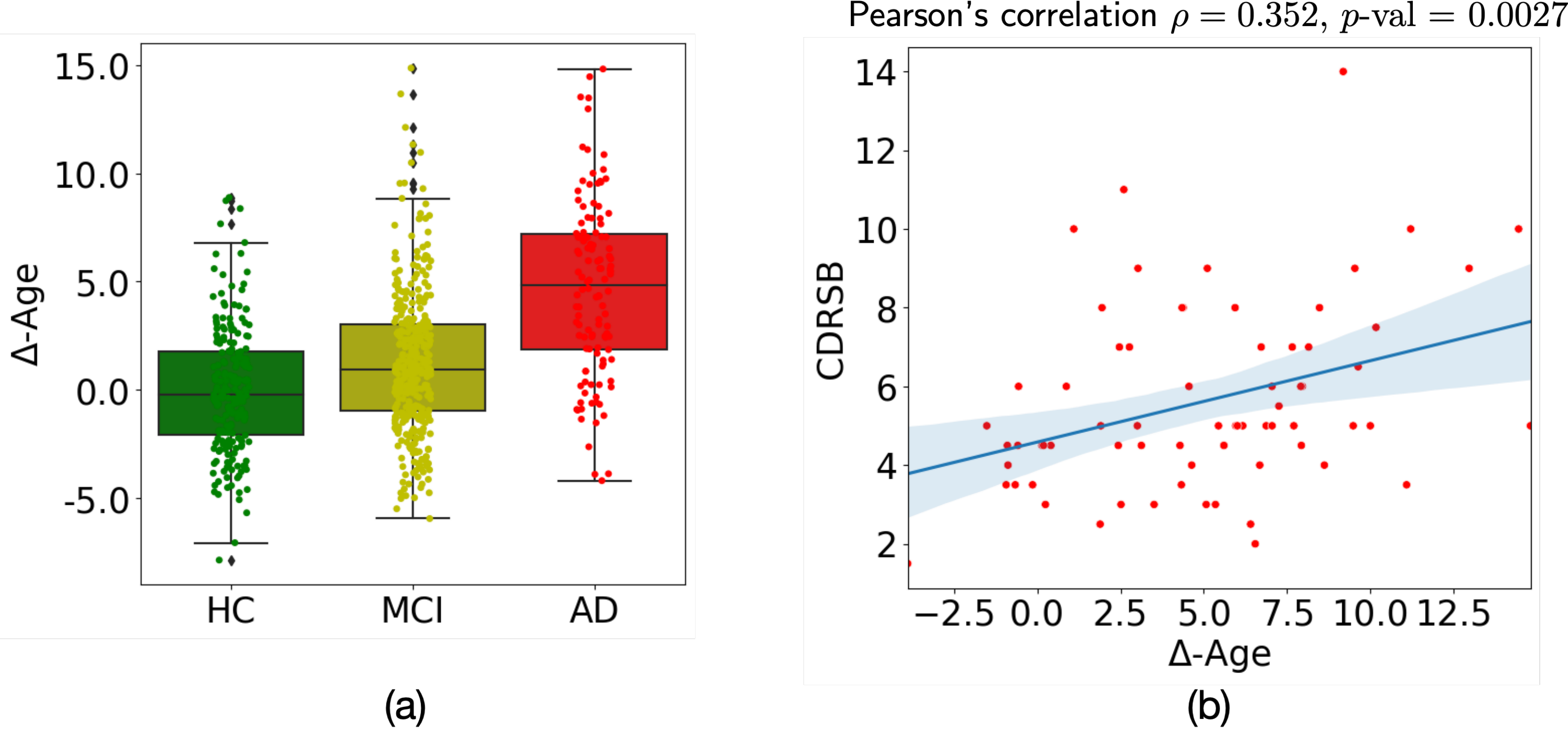}
	\caption{\footnotesize (a) Distributions of $\Delta$-Age in HC, MCI, and AD cohorts. (b) Scatterplot between the CDRSB and $\Delta$-Age for the AD cohort. The line representing the linear fit is also shown.}	\vspace{20pt}
\end{minipage}

Figure~3 \textcolor{black}{reproduces the results from~\cite{sihag2024brain}} and validates that $\Delta$-Age inferred by the GNN model is a biomarker of AD. In fact, the $\Delta$-Age distribution has the highest mean for the AD group ($\Delta$-Age = 4.66 $\pm$ 4.04 years) and the smallest mean for the HC group ($\Delta$-Age = 0 $\pm$ 2.9 years); the MCI cohort lies in between them ($\Delta$-Age = $1.23\pm 3.29$ years). Furthermore, $\Delta$-Age in the AD group was significantly correlated with CDRSB (Pearson's correlation = $0.352$, $p$-val = 0.0027).
\end{mdframed}

The studies in~\cite{franke2010estimating, lowe2016effect, yin2023anatomically} primarily considered AD as the disease group and consistently reported elevated $\Delta$-Age in AD relative to the healthy population. Moreover, the study in~\cite{lowe2016effect} stratified the population according to APOE $\varepsilon_4$ carriers and non-carriers (a gene variant that increases the risk of developing AD) to report varying $\Delta$-Age in these subgroups, which also exhibited different patterns of longitudinal progression of the disease. The focus of~\cite{yin2023anatomically} was to demonstrate the anatomical interpretability of brain age in AD and different subgroups within the healthy population, through prevalent methods from explainable artificial intelligence (AI). Similar analyses were adopted in other disease-specific $\Delta$-Age studies for schizophrenia~\cite{schnack2016accelerated}, TBI~\cite{cole2015prediction}, and Parkinson's disease~\cite{eickhoff2021advanced}. 

\begin{myboxii}
    \textcolor{black}{As pre-trained models, brain age gap prediction algorithms are widely applicable to derive biomarkers for numerous clinically-defined health conditions associated with neurodegeneration.}
\end{myboxii}

A common theme in all the aforementioned studies is that the brain age gap prediction model was trained only on a healthy population, and then deployed to study $\Delta$-Age for specific neurodegenerative conditions. Hence, the ML workflow for $\Delta$-Age prediction has a generalist characteristic as it is not bound to a specific health condition, unlike many supervised learning algorithms developed to study neurodegeneration. This key observation has been well documented~\cite{yin2023anatomically,eickhoff2021advanced,sihagisbi}, as multiple studies derived anatomical patterns to confirm that disease-relevant features contributed to the reported brain age (or $\Delta$-Age). Higher $\Delta$-Age was found for AD, FTD, and LBD in~\cite{eickhoff2021advanced}, followed by distinct anatomical characterizations of these neurodegenerative conditions using explainable AI tools. In a similar spirit, the study in~\cite{sihagisbi} leveraged a GSP-driven explainable model for $\Delta$-Age prediction~\cite{sihag2023explainable} to report distinct anatomical characterizations of $\Delta$-Age in AD, FTD, and the combined group of CBS and PSP conditions. \textcolor{black}{Existing studies have also shown the effectiveness of brain age gap in other clinical domains, such as for major depressive disorder~\cite{ho2024atypical} and to track changes throughout the human lifespan~\cite{bethlehem2022brain}. For a comprehensive review of  brain age gap prediction algorithms applied to different biological domains, the interested reader is referred to~\cite{franke2019ten}. Moreover, brain age gap has been discussed within the broader context of biological age estimation~\cite{cole2019brain}. In summary, the body of work surveyed in this section justifies the adoption of brain age gap as a promising biomarker for the early stages of clinical workflows, where individuals may not yet have a clear diagnosis.}





\begin{table}
\caption{A summary of the studies on $\Delta$-Age for various neurodegenerative conditions and associated experiments that validate $\Delta$-Age as a biomarker.}
    \centering
    \begin{tabular}{|c|c|c|}
    \hline
        {\bf Clinical Condition} & {\bf Experiments establishing $\Delta$-Age as a Biomarker} & {\bf Reference}\\
        \hline
        AD & Elevated $\Delta$-Age in disease group & ~\cite{franke2010estimating}\\
        \hline
        \multirow{2}{*}{AD} & Differentiating $\Delta$-Age in APOE $\varepsilon_4$ carriers/non-carriers, & \multirow{2}{*}{\cite{lowe2016effect}}\\
        & longitudinal progression, correlation with neuropsychological test scores &\\
        \hline
         \multirow{2}{*}{AD} & Elevated $\Delta$-Age in disease group, anatomical maps, & \multirow{2}{*}{\cite{yin2023anatomically}}\\
        & associations with neurocognitive measures  &\\
        \hline
     {AD, Frontotemporal } & Elevated $\Delta$-Age in disease groups, distinct anatomical maps for $\Delta$-Age, & \multirow{3}{*}{\cite{lee2022deep}}\\
       Dementia (FTD), & associations with neurocognitive measures, tau PET  &\\
       Lewy Body Dementia (LBD) & & \\
        \hline
Schizophrenia & Elevated $\Delta$-Age in disease groups, longitudinal analysis & \cite{schnack2016accelerated}\\
\hline
Traumatic Brain Injury (TBI) & Elevated $\Delta$-Age in TBI, correlation with time since injury  & \cite{cole2015prediction}\\
\hline
\multirow{2}{*}{Parkinson's Disease (PD)}  & Elevated $\Delta$-Age in disease group, associations with  & \multirow{2}{*}{~\cite{eickhoff2021advanced}}\\
& cognitive and motor impairment & \\
\hline
        AD, FTD, & Elevated $\Delta$-Age in disease groups, & \multirow{3}{*}{\cite{sihagisbi}}\\
        Corticobasal Syndrome (CBS), & distinct anatomical maps for $\Delta$-Age in disease groups & \\
        Progressive Supranuclear Palsy (PSP) & &\\
        \hline
    \end{tabular}
    
    \label{tab_review}
\end{table}

\subsection*{Performance-driven approach to brain age gap prediction: Unresolved challenges and gaps}

Despite promising results in characterizing neurodegeneration, there exists considerable divergence in the underlying ML principles adopted for $\Delta$-Age prediction~\cite{gaser2024perspective}. The diversity of \textcolor{black}{ML models} chosen for this application notwithstanding (see e.g.,~\cite{azzam2024review} for a recent review), here we elucidate the fundamental methodological obscurities in $\Delta$-Age prediction by performance-driven principles that severely challenge their adoption in practice. These limitations have been well documented in the literature~\cite{bashyam2020mri, jirsaraie2023systematic,sihag2023explainable}. 

In light of the challenge stemming from the lack of a tangible ground truth for brain age, the focus of most ML-driven approaches in this domain has overwhelmingly been on: (i) achieving near perfect performance on the chronological age prediction task for healthy individuals; and (ii) using it as a measure to gauge the quality of a $\Delta$-Age prediction framework. Here, we use the taxonomy `performance-driven approach' to $\Delta$-Age prediction for such ML methods. Performance-driven approaches gauge their quality through metrics such as mean absolute error (MAE) on chronological age prediction in healthy populations. They encompass both traditional ML-driven methods and deep learning models, and the higher expressive power of the latter makes them the prevalent choice today~\cite{azzam2024review}. 

In principle, achieving the best possible performance in predicting chronological age in a healthy population is well motivated, as it allows the ML model to learn patterns of healthy aging. However, as is abundantly clear from Table~\ref{tab_review} and our prior discussion in `Brain age gap as a biomarker of neurodegeneration', the \emph{validation} of $\Delta$-Age as a biomarker of health conditions is critical for its meaningful adoption in clinical settings. 

\begin{myboxii}
    For $\Delta$-Age to be a valid biomarker, it is unclear how accurate the underlying ML brain age gap prediction algorithm must be when used to predict chronological age for a healthy population.
\end{myboxii}

There exists ample evidence in the literature to corroborate that a more accurate prediction of chronological age in healthy populations does not necessarily translate into improved validation of the inferred $\Delta$-Age as a biomarker. The study in~\cite{bashyam2020mri} reported that a $\Delta$-age prediction model with a relatively `moderate' fit on the chronological age of healthy individuals led to inferred $\Delta$-Age with better clinical utility in neurodegenerative conditions, when compared to a model that had a `tighter' fit on the chronological age of the same healthy cohort. Similar findings were reported in~\cite{sihag2023explainable}, where the model with higher MAE on chronological age predictions of the healthy population inferred $\Delta$-Age with a higher correlation with CDRSB scores in AD (relative to a baseline model with smaller MAE). A comprehensive evaluation of various ML approaches in~\cite{jirsaraie2023systematic} found no significant correlation between the accuracy of chronological age prediction and the clinical utility of the accompanying $\Delta$-Age estimates. \textcolor{black}{Figure~4 summarizes the methodological obscurity in $\Delta$-Age prediction by performance-driven approaches.}  

\begin{figure}
    \centering
    \includegraphics[width=\linewidth]{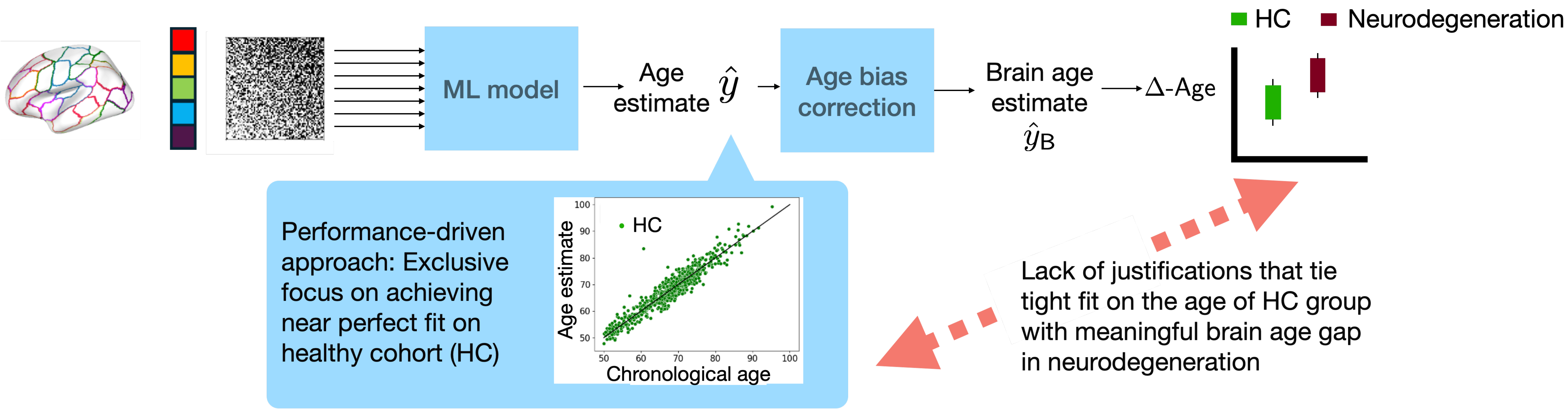}    \caption{\textcolor{black}{Performance-driven approaches to $\Delta$-Age prediction prioritize a near-perfect fit on the chronological age of the HC, yet they lack the conceptual or statistical justifications to ensure the relevance of inferred $\Delta$-Age as a biomarker for neurodegeneration.}}
    \label{fig:perf_driven}
\end{figure}

Chronological age clearly acts as a proxy for biological age in healthy individuals when pre-training an ML model for $\Delta$-Age prediction. However, the discussion above \emph{does not} lead to the conclusion that pre-training the model on a healthy cohort for chronological age prediction is inherently flawed. A more nuanced observation to make is that the overwhelming focus on achieving a near-perfect fit on the chronological age of the healthy population potentially overlooks meaningful
biological information necessary to discriminate between the healthy
and clinical groups~\cite{gaser2024perspective}. 
We contend that this could be attributable to performance-driven approaches' neglect of the heterogeneity of biological aging within the healthy population itself, due to factors such as obesity, stress levels, among others. These variables also affect the propensity to develop a neurodegenerative condition in the future, within the healthy population.

\begin{myboxii}
    Performance-driven approaches tend to take the leap from achieving the `best possible fit' with a principled ML model to a `near perfect fit' on chronological age by arbitrary additions of sophisticated computation modules, compounding methodological obscurities in $\Delta$-Age prediction.
\end{myboxii}

Recent performance-driven studies have adopted model-agnostic, post-hoc methods from explainable AI to corroborate the biological validity of their brain age predictions. For instance, saliency maps were utilized in~\cite{yin2023anatomically} and~\cite{lee2022deep} to identify those brain regions most relevant to predicted brain age for different clinical groups. However, such approaches still provide an incomplete perspective to $\Delta$-Age as a biomarker. Specifically, since $\Delta$-Age is a deviation from the chronological age, an explanation of brain age alone overlooks the component of the deviations due to healthy brain aging. Additionally, the explanations offered by post-hoc explainability methods suffer from lack of robustness, for example, due to instability to small perturbations in the input, variability in explanations due to stochasticity in training algorithms, and model multiplicity (i.e., when multiple models with similar performance may exist but offer distinct explanations)~\cite{dombrowski2019explanations,adebayo2018sanity,black2022model}. Therefore, further progress is needed within the explainable AI domain for these approaches to be confidently adopted in practice. 

\begin{mdframed}[hidealllines=true,backgroundcolor=gray!20]
{\bf Adding mathematical depth to brain age gap prediction.}
It is unlikely that the conceptual gap in performance-driven approaches regarding the statistical dependency between the accuracy of the model during training on healthy individuals and $\Delta$-Age as a biomarker for neurodegeneration can be bridged with experiments alone. Therefore, the development of relevant mathematical principles is critical for a viable and generalizable practical methodology. To this end, we identify the following four mathematical principles:

\begin{itemize}
	\item {\bf Principle 1: Focusing on $\Delta$-Age as a residual of regression tasks.} The residuals of the regression models inform the $\Delta$-Age estimates when the ML model is deployed to predict chronological age for individuals with adverse health conditions. Hence, it is paramount to focus on the structure and statistics of the residuals of the age prediction model, rather than the age prediction task itself, to validate the viability of $\Delta$-Age as a biomarker. 
	
	\item {\bf Principle 2: Shift focus to qualitative evaluation of ML models trained on the healthy population.}  It is key to go beyond the performance on the chronological age prediction task and instead focus on a holistic characterization of the representations that the ML model learns when exposed to data from the healthy population. 
	
	\item {\bf Principle 3: Transference of pre-trained age prediction models to neurodegenerative cohorts for constructing $\Delta$-Age as a biomarker.}  In principle, predicting $\Delta$-Age in neurodegenerative cohorts can be perceived as an unsupervised transfer learning problem, where we expect a degradation in performance. This problem is unsupervised because the expected amount of drift in model performance is unknown (i.e., there is no ground truth for brain age in neurodegenerative cohorts). It would be desirable to identify the specific deviations in the information processing pipeline itself, which are the contributors to the elevated $\Delta$-Age observed in neurodegeneration.
	
	\item {\bf Principle 4: Generalizability beyond specific dimensionality of the data.} Due to the existence of different brain atlases, the neuroimaging datasets characterizing the same population can have arbitrary dimensionalities in independently conducted studies~\cite{woolsey2017brain}. This creates a challenge for reproducibility of findings, as the prevalent performance-driven approaches in both traditional ML and deep learning are limited within the dimensionality of the dataset that they have been trained on. To address this challenge, we need mathematical principles that govern the reproducibility of findings derived using an ML model across datasets of different dimensionalities, without being re-trained from scratch. 	
\end{itemize}
\end{mdframed}

\begin{myboxii}
    Adoption of Principles 1-4 in brain age gap prediction will fundamentally shift the primary focus to $\Delta$-Age prediction as a biomarker, rather than being a byproduct of age prediction in performance-driven approaches.
\end{myboxii}

Our discussion on `Adding mathematical depth to brain age gap prediction' summarized desirable mathematical principles behind brain age gap prediction to improve its practical viability. \textcolor{black}{Traditional ML methods or the prevalent deep learning models adhere to some, but not all the principles identified above. For instance, a PCA-regression model could address the requirements of Principle 2 via transparent evaluation of 
$\Delta$-Age, but at the same time, such a model may suffer from instability~\cite{sihag2022covariance}. On the other end of the spectrum, deep learning models can offer improved robustness and advanced operational capabilities, thus meeting the requirements of Principle 4. However, these `black-box' architectures output predictions that are not inherently explainable; thus, they fall short in meeting the requirements of Principles 1-3. For instance, CNNs are commonly adopted in the brain age gap prediction pipeline~\cite{bashyam2020mri, yin2023anatomically}, but fail to convincingly reveal the anatomical factors contributing to brain age gap in neurodegeneration. 
Based on the preceding discussion, we argue that the alignment of Principles 1-4 with the brain age gap prediction pipeline is nontrivial and requires a deeper \emph{theoretical} understanding of the chosen ML model. GSP offers an ideal suite of foundational tools for structured multivariate information processing and ML over graphs, facilitating the desired explainability, robustness, and transferability analyses.}    



\section*{GSP Foundations for Neuroimaging Data Analysis}

Recent GSP advances have led to principled and theoretically sound learning tools for a variety of applications where data reside in non-Euclidean domains and exhibit graph structure~\cite{leus2023graph}. GSP  paved the way for innovative GNN architectures
, thus, bridging signal processing insights and mathematical theory with the empirical successes of deep learning~\cite{ruiz2021graph}. The domain of network neuroscience, which studies the brain via its network representations, has been a major beneficiary of these advances in GSP due to a concurrent increase in the availability of large spatiotemporal MRI datasets~\cite{bassett2017network, huang2018graph}. 

\begin{myboxii}
    GSP offers a natural substrate over which the mathematical advancements needed for a practically viable brain age gap prediction workflow can be developed. 
\end{myboxii}

\subsection*{GSP and GNNs: An overview}

\textcolor{black}{We review the background on GSP and GNNs needed to introduce an explainable brain age gap prediction framework adhering to Principles 1-4; see also~\cite{leus2023graph, ruiz2021graph} for other insightful tutorial treatments.} 

The standard information processing backbone in GSP can be described in terms of four main pillars. 
First, consider a \emph{graph} ${\cal G}({\cal V}, {\cal E}, {\cal W})$ with a set of $M$ nodes ${\cal V} = \{1,\dots, M\} $, a set of undirected edges ${\cal E} \subseteq {\cal V}\times {\cal V} $, and an edge weight function ${\cal W}: {\cal E}\mapsto \mathbb{R}$. The graph topology can be compactly represented using a symmetric matrix ${\bf A}$ of size $M\times M$, which encodes the edge and weight information. The adjacency and Laplacian matrices are examples of such commonly used matrix representations. \textcolor{black}{In neuroimaging data analysis, the graph is typically data-driven, with its nodes representing different brain regions and the edges and weights inferred from nodal data; e.g., pairwise correlations among cortical thickness features (Fig. \ref{fig:smrifts}). 
}
Second, a \emph{graph signal} ${\bf x} = [x_1,\dots, x_M]^\top\in\mathbb{R}^M$ is a vector representation of the data supported on the graph ${\cal G}$, i.e., each element within ${\bf x}$ can be associated with a distinct node in ${\cal V}$. \textcolor{black}{Going back to Case Study 1, the vector of cortical thickness features for an individual represents their graph signal. }
Third, a \emph{graph filter} is the computational module to transform the graph signal ${\bf x}$ over the graph representation ${\bf A}$~\cite{isufi2024graph}. Information processing with a graph filter relies on the \emph{shift} operation ${\bf A}{\bf x}$, which shifts the graph signal $\mathbf{x}$ over the nodes in ${\bf A}$, such that, the $i$-th component of ${\bf Ax}$ is
    \begin{align}
        [{\bf A}{\bf x}]_i = \sum\limits_{j = 1}^M [{\bf A}]_{ij} {\bf x}_j,
    \end{align}
i.e., its value is determined by an aggregation of the information in ${\bf x}$ according to the weights in the $i$-th row of ${\bf A}$ (corresponding to edges incident to node $i$). In general, the output of the shift operation, ${\bf Ax}$, is another graph signal, whose elements are obtained by \textcolor{black}{linear mixing of the elements in }${\bf x}$ according to the weights in ${\bf A}$. Building upon this observation, the graph filter implements the convolution operation via a polynomial on ${\bf A}$, such that, the output of the graph filter is
    \begin{align}\label{gf}
       {\bf z} = {\bf H}({\bf A}) {\bf x}, \quad \text{where }\: {\bf H}({\bf A}) =  \sum\limits_{k = 0}^{K}h_k {\bf A}^k ,
    \end{align}
and ${\bf h} = [h_0, h_1,\dots, h_{K}]^\top \in\mathbb{R}^{K}$ are the \emph{filter taps}.  
%
\textcolor{black}{Finally, the \textit{graph Fourier transform} (GFT) 
facilitates a spectral decomposition of graph signals and filters.
Consider the eigendecomposition ${\bf A} = {\bf U}{\bf \Lambda} {\bf U}^{\top}$, where ${\bf U} = [{\bf u}_1,\dots, {\bf u}_M]$ is the orthonormal matrix of $M$ eigenvectors, and ${\bf \Lambda} = {\sf diag}(\lambda_1,\dots, \lambda_M)$ is the diagonal matrix of eigenvalues ordered as $\lambda_1\geq \lambda_2\geq \dots\geq \lambda_M$. The graph Fourier transform (GFT) is defined as the projection of the signal ${\bf x}$ onto the eigenspace of ${\bf A}$, namely}
\begin{align}\label{gft}
    \tilde {\bf x} = {\bf U}^{\top}{\bf x}\;.
\end{align}
\textcolor{black}{The $i$-th entry of $ \tilde {\bf x}$ quantifies the contribution of \textcolor{black}{${\bf u}_i$} to the graph signal ${\bf x}$ via the inner product $[\tilde{\bf x}]_i={\bf u}_i^{\top} {\bf x}$. Indeed, from \eqref{gft} it follows that a graph signal can be synthesized as $\mathbf{x}=\sum_{i=1}^M [\tilde{\bf x}]_i\mathbf{u}_i$. Eigenvector ${\bf u}_i$ is the frequency component associated to frequency $\lambda_i$, and $[\tilde{\bf x}]_i$ the corresponding GFT coefficient of $\mathbf{x}$.}

%
%
%
%
Taking the GFT of the graph filter output in~\eqref{gf} and considering the eigendecomposition of ${\bf A}$ yields
\begin{align}\label{gftgf}
    \tilde {\bf z} &= {\bf U}^{\top}{\bf z} = {\bf U}^{\top} \sum\limits_{k = 0}^{K} h_k{\bf U}{\bf \Lambda}^k {\bf U}^{\top} {\bf x} = \sum\limits_{k = 0}^{K} h_k {\bf \Lambda}^k {\bf U}^{\top} {\bf x} = {\bf H}({\bf \Lambda})\tilde {\bf x},
\end{align}
where ${\bf H}({\bf \Lambda}) = {\sf diag}(h(\lambda_1), \dots, h(\lambda_M))$ is a diagonal matrix and 
\begin{align}\label{frere}
    h(\lambda_i) := \sum\limits_{k = 0}^{K} h_k \lambda_i^k\;.
\end{align}

%

\noindent
By inspection of~\eqref{gft}-\eqref{frere}, it follows that the impact of the graph filter on the $i$-th element of $\tilde{\bf x}$ (the inner product ${[\tilde{\bf x}]_i=\bf u}_i^{\top} {\bf x}$) is limited to a scaling by  
$h(\lambda_i)$, \textcolor{black}{in a way akin to a convolution theorem}, i.e.,
\begin{align}\label{eq9}
    [\tilde{\bf z}]_i = \textcolor{black}{h(\lambda_i) [\tilde{\mathbf{x}}]_i} = h(\lambda_i) {\bf u}_i^{\top}{\bf x}\;.
\end{align}
Therefore, the graph filter modifies the contribution of the $i$-th component ${\bf u}_i$ to the output via the function $h: \mathbb{R}\mapsto \mathbb{R}$ evaluated on the eigenvalue $\lambda_i$. Accordingly, $h(\lambda_i)$ is known as the \emph{frequency response} of the graph filter ${\bf H}(\cdot)$ at frequency $\lambda_i$~\cite{isufi2024graph}.
In supervised ML settings, the filter taps form the~\emph{learnable} parameters that are estimated from data. \textcolor{black}{To reconcile these GSP concepts with the brain age gap prediction task at hand, we note that graph filter taps (and the resulting frequency response) can qualitatively capture the impact of the training procedure, thus, making Principle 2 actionable; see also `Enter VNNs: GNNs with covariance graphs for structural MRI'.} While the capacity of a graph filter is limited to learning \emph{linear} operators, they are the key ingredients in GNNs -- the subject dealt with next.

\noindent {\bf Graph neural networks.} GNNs are learnable parametric architectures for \emph{nonlinear} information processing, which are built from graph filter primitives. A \textit{graph perceptron} is constructed by feeding the output of the graph filter through a  pointwise nonlinear activation function $\sigma(\cdot)$ (e.g., ${\sf ReLU}, \tanh$), that satisfies \textcolor{black}{$\sigma({\bf d}) = [\sigma(d_1), \dots, \sigma(d_M)]^\top$ for ${\bf d} = [d_1, \dots, d_M]^\top$}. Hence, the output of a single layer GNN with input $\bx$ is given by $\bz = \sigma(\bH(\bA) \bx)$. For an $L$-layer GNN, let $\bH_{\ell}(\bA)$ be the graph filter in layer $\ell$ and $\cH_{\ell}$ the corresponding set of filter taps. A multilayer (deep) GNN can simply be formed by concatenating individual graph perceptrons, such that the recursive relationship between the input $\bx_{\ell-1}$ and the output $\bx_{\ell}$ at the $\ell$-th layer is $\bx_{\ell} = \sigma(\bH_{\ell}(\bA)\bx_{\ell-1}),\text{ for } \ell\in \{1,\dots,L\}$, where $\bx_0$ is the input $\bx$. 

The expressive power of a GNN architecture can be further enhanced by incorporating multiple input multiple output (MIMO) processing at every layer\textcolor{black}{; see~\cite{ruiz2021graph} for additional details we omit here to avoid introducing unnecessarily cumbersome notation.} 
In any case, a multilayer GNN architecture capable of MIMO processing is henceforth denoted as $\Psi(\bx;\bA,{\cal H})$, where the set of filter taps $\cH$ captures the full span of its architecture. We also write $\Psi(\bx;\bA,{\cal H})$ to denote the output at the GNN's final layer, which is the GNN representation of input $\bx$. The output $\Psi(\bx;\bA,{\cal H})$ is typically succeeded by a readout function that maps it to the desired inference outcome.

\begin{myboxii}
    The theoretical and operational properties that make GSP appealing to neuroimaging data analysis begin to take shape at the level of the graph filter itself.
\end{myboxii}
\noindent {\bf Stability of graph filters and GNNs.} Interestingly, graph filter outputs are \emph{stable} to various abstract perturbations of~${\bf A}$. More formally, $\| {\bf H}({\bf A}) - {\bf H}(\bA + \delta{\bf A})\|$ is provably bounded for a controlled $\delta \bA$, provided the frequency response $h(\lambda)$ is sufficiently smooth (so-termed Lipschitz conditions)~\cite{ruiz2021graph}. This property suggests that ML models using graph filters may provide reproducible outcomes, \textcolor{black}{which is relevant to reproducibility in network neuroscience when brain graphs are estimated from different datasets (or sample sizes).} Notably, the stability of graph filters readily extends to GNNs; $\| \Psi(\bx; \bA, \cH) - \Psi(\bx; \bA + \delta\bA, \cH)\|$ is bounded under similar mild Lipschitz conditions on the constituent graph filters. 

\noindent {\bf Transferability of graph filters and GNNs.} A given graph filter $\bH(\cdot)$, i.e., \textcolor{black}{with fixed filter taps}, can be \emph{transferred} to process datasets of arbitrary dimensionalities. \textcolor{black}{Indeed, the same polynomial function $\bH(\cdot)$ can be evaluated on matrices of any size.} Consider another \textcolor{black}{$M'$}-dimensional graph signal ${\bf x}' = [x'_1,x'_2,\dots, x'_{M'}]^\top$ associated with a graph representation ${\bf A}'$ of size \textcolor{black}{$M'\times M'$}. In this scenario,  the filter taps ${\bf h}$ in~\eqref{gf} can be reused to generate another output ${\bf z}' = \sum_{k = 0}^{K}h_k {\bf A}'^k{\bf x}' = {\bf H}({\bf A'}){\bf x}'$, now a vector of length $M'$. This property readily extends to GNNs based on graph filters, where the same GNN can be used to generate nodal representations from different datasets of distinct dimensionalities.
\textcolor{black}{This transferability property supports Principle 4 on `Generalizability beyond specific dimensionality of data'. Granted, the success of transference will be measured by the consistency in performance attained across datasets curated according to different brain atlases. Hence, studying the convergence between the graph filter outputs $\mathbf{z}$ and $\mathbf{z}'$ will be central to the theoretical characterization of successful transference.} 

The convergence between $\mathbf{z}$ and $\mathbf{z}'$ can be formalized by considering the continuous approximations of these discrete objects. Specifically, given an $M$-dimensional vector $\bx = [x_1,\dots,x_M]^\top$, we can define a continuous representation of $\bx$ as a function $y_{\bx}: [0,1] \mapsto \mR$, such that, \textcolor{black}{$y_{\bx}(a) = x_i$ for $a\in \cU_i$}, where $\cU_i$ is a pre-defined subinterval of $[0,1]$ associated with the $i$-th element of $\bx$. Similarly, we can map the matrix $\bA$ to a compact set $[0,1]^2$ via $W_{\bA}: [0,1]^2 \mapsto \mR$, where we have \textcolor{black}{$W_{\bA}(a,b) = [\bA]_{ij}$ for $a\in \cU_i$ and $b\in \cU_j$}. 
Note that we can recover $\bx$ from $y_{\bx}$ and vice-versa (similarly for $\bA$ and $W_{\bA}$). Hence, for graph signals $\bx$ and $\bx'$ consisting of $M$ and $M'$ elements, respectively, the closeness of their continuous representations $y_{\bx}$ and $y_{\bx'}$, i.e., $\|y_{\bx} - y_{\bx'}\|$  can be used to quantify the similarity between graph signals of different lengths. This observation also extends to the comparison between matrices $\bA$ and $\bA'$. By leveraging the theory of graphons as limit objects of graphs (i.e., when $M\rightarrow \infty$)~\cite{lovasz2012large}, the convergence between filter outputs ${\bf z}$ and ${\bf z}'$ via their respective continuous approximations $y_{\bf z}$ and $y_{{\bf z}'}$ was established using so-termed graphon signal processing~\cite{ruiz2021graph}. Specifically, under \textcolor{black}{smoothness conditions on the graph filter (i.e., the variation between the frequency responses $h(\lambda_i)$ and $h(\lambda_j)$ is bounded as $|h(\lambda_i) - h(\lambda_j)| \leq \vartheta |\lambda_i - \lambda_j|$ for some $\vartheta > 0$ and any pair $(\lambda_i, \lambda_j)$)} and the assumption that the continuous approximations $W_{\bf A}$ and $W_{\bf A'}$ are part of a converging sequence, the distance $\|y_{\bf z} - y_{{\bf z}'}\|$ vanishes at the rate of $\frac{1}{\sqrt{M}} + \frac{1}{\sqrt{M'}}$ for a graph filter instantiated on graphs $\bA$ and $\bA'$ of sizes $M$ and $M'$, respectively.

In summary, our GSP-friendly exposition of GNNs revealed that learned representations are intimately tied to the graph eigenvectors, thus, lending some notion of explainability to information processing with GNNs. Moreover, the stability and transferability properties of GNNs support their generalizability to diverse settings. It is precisely in this sense that GSP provides the appropriate substrate to develop the necessary mathematical principles identified in `Adding mathematical depth to brain age gap prediction'.  

\subsection*{Enter VNNs: GNNs with covariance graphs for structural MRI}

Data-driven graphs are ubiquitous in neuroimaging data analysis. The anatomical covariance matrix estimated from features derived from structural MRI is a prominent example~\cite{evans2013networks}. \textcolor{black}{Noteworthy contributions on morphometric similarity networks have generalized anatomical covariances to include multiple information modalities within structural MRI~\cite{seidlitz2018morphometric}, with impact to brain age gap prediction~\cite{galdi2020neonatal}.}

There are critical theoretical gaps in the existing theoretical and empirical properties of GNNs outlined in `GSP and GNNs: An overview', which are agnostic to the nuanced spatial and spectral characteristics inherent to a data-driven graph. 
\textcolor{black}{In particular, successful practical adoption of GNNs instantiated on covariance matrices is contingent on a refined mathematical understanding of their properties. In this section, we bridge this gap by surveying the mathematical foundations of data analysis when data-driven covariance graphs are used in GNNs. These architectures are known as coVariance neural networks (VNNs)~\cite{sihag2022covariance, cavallo2024sparsecovarianceneuralnetworks, cavallo2024spatiotemporal}, and we use the terminology `coVariance filter' to refer to a graph filter implemented on a covariance matrix. Importantly, we discuss the implications of these results on the brain age prediction task deal with here.}

\begin{figure}[t]
    \centering
    \includegraphics[width=0.9\linewidth]{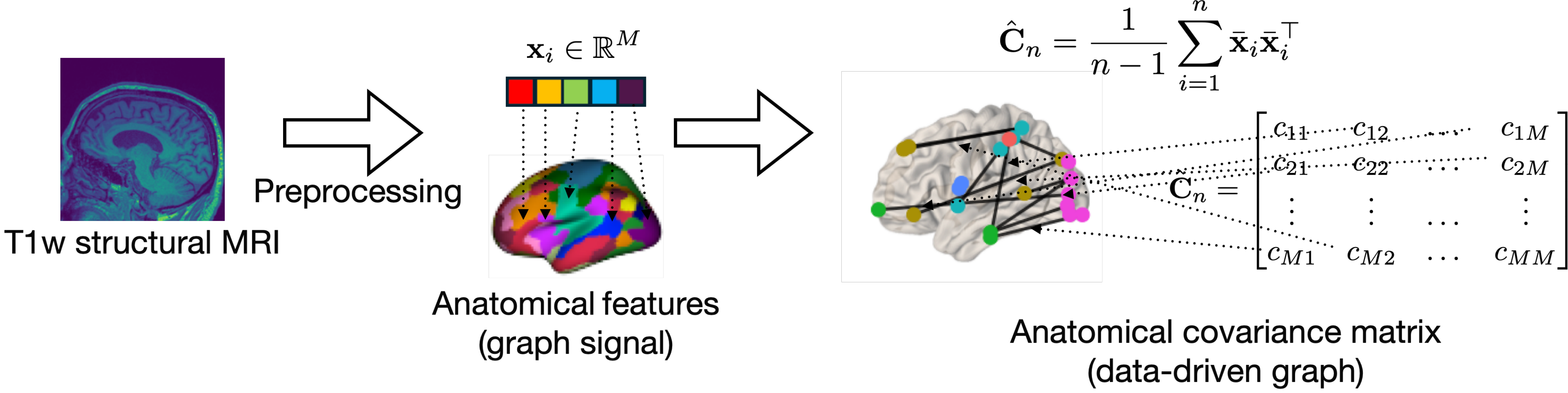}
    \caption{Extracting graph signals and a data-driven graph from structural MRI. Pre-processing of structural MRI yields anatomical features across the brain cortex. These anatomical features have a vector representation ${\bx}$ of length $M$, each element of which corresponds to a distinct brain region. From a dataset of anatomical features from $n$ individuals, we can estimate the anatomical covariance matrix $\hat \bC_n \in \mathbb{R}^{M\times M}$. \textcolor{black}{Here, the shorthand notation $\bar \bx_i$ stands for $(\bx_i - \bar\bx)$ in~\eqref{sample_cov}.} The anatomical covariance matrix comprises the graph representation of brain anatomy, with its off-diagonal elements characterizing the correlation between anatomical features associated with different brain regions.}
    \label{fig:smrifts}
\end{figure}

\noindent
{\bf Covariance matrix.} Covariance matrices are fundamental data structures within multivariate data analysis, that encode statistical dependencies between different pairs of features in a dataset. Our perspective is to view a covariance matrix as a graph representation of a multivariate dataset consisting of $n$ random, independent and identically distributed (i.i.d.) data samples~$\bx_i \in \mR^{M },\forall i\in\{1,\dots,n\}$. \textcolor{black}{In our neuroimaging setting, the data samples are the anatomical features, where $M$ is the number of brain regions of interest (Fig. 5).} The empirical covariance matrix is estimated from samples as
\begin{align}\label{sample_cov}
    \hat \bC_{n} \triangleq \frac{1}{n-1} \sum\limits_{i=1}^n(\bx_i - \bar\bx) (\bx_i-\bar\bx)^{\top}\;,
\end{align}
%
where $\bar \bx$ is the sample mean across $n$ samples. \textcolor{black}{When samples correspond to cortical brain features, the anatomical covariance matrix is the mathematical construct encoding pairwise statistical interdependencies of brain atrophy across brain regions. Figure~\ref{fig:smrifts} illustrates the anatomical features obtained from structural MRI scans (the graph signals) and the process of estimating the anatomical covariance matrix that is used as a graph representation of brain anatomy.} From a statistical perspective, the sample covariance matrix is a statistical estimate of the true covariance matrix (also referred to as ensemble covariance matrix) ${\bf C}$. This ensemble covariance matrix $\bC$ is determined from an $M-$dimensional random vector $\bx \in \mR^{M}$ as $\bC \triangleq \mE[(\bX-\mE[\bx])(\bx-\mE[\bx])^{\top}]$.  Although the graph filters and GNNs can be analogously defined on both $\bC$ and $\hat\bC_n$, the ensemble covariance matrix $\bC$ cannot be observed directly. \textcolor{black}{Thus, in practice we use noisy sample-based statistical estimates and their quality relative to the ensemble counterparts is governed by matrix perturbation theory~\cite{loukas2017close}}. This observation also extends to the eigenspectrum of $\hat\bC_n$ and $\bC$. \textcolor{black}{Specifically, consider the eigendecomposition $\bC = \bV \boldsymbol{\Phi} \bV^{\top}$, where $\bV  = [\bv_1,\dots, \bv_M]$ is the $M\times M$ matrix of eigenvectors and $\boldsymbol{\Phi} = {\sf diag}(\phi_1,\dots,\phi_M)$ is the diagonal matrix of eigenvalues $ \phi_1\geq \phi_2\geq \dots \geq \phi_M\geq 0$. Then, the eigendecomposition of the sample covariance matrix is $\hat\bC_n = \hat\bV \hat{\boldsymbol \Phi} \hat\bV^{\top}$, where its matrix of eigenvectors $\hat\bV = [\hat\bv_1,\dots, \hat\bv_M]$ are statistical estimates of $\bV$ and the eigenvalues $\hat{\boldsymbol \Phi} = {\sf diag}(\hat\phi_1,\dots, \hat\phi_M)$ are statistical estimates of $\boldsymbol \Phi$. Matrix $\hat{\boldsymbol \Phi}$ is a perturbed version of ${\boldsymbol \Phi}$.}
 
Case study 1 demonstrated that a reduction in cortical thickness is characteristic of brain atrophy, which manifests in both healthy aging and neurodegeneration. This implies that the correlation structure in the anatomical covariance matrix will be distorted when specific brain regions exhibit accelerated brain atrophy due to neurodegeneration, relative to that of the healthy cohort. Based on this discussion, we contend that a GNN with anatomical features as inputs and the anatomical covariance matrix as the graph provides a suitable framework for the $\Delta$-Age prediction pipeline. These VNN models are discussed next.

\noindent{\bf CoVariance neural networks and links with PCA.} 
The eigenspectrum of the covariance matrix implicitly captures the structure of a dataset via the \textit{principal components}, and said structure can be exploited via the PCA transform~\cite{shlens2014tutorial}. PCA-regression has been integrated into brain age gap prediction pipelines dating back to the first studies in the field~\cite{cole2015prediction}. \textcolor{black}{Interestingly, a coVariance filter draws similarities with the PCA transform~\cite[Theorem 1]{sihag2022covariance}. This connection follows directly from~\eqref{gftgf} and~\eqref{frere}. Specifically, the output of the graph filter instantiated on $\hat\bC_n$ (i.e., a coVariance filter) depends on the projection of anatomical features onto the covariance eigenvectors -- the principal components $\hat{\bf v}_i^{\top} {\bf x}$.}

\textcolor{black}{When it comes to qualitative assessment for neuroimaging data analysis, the equivalence between PCA and coVariance filters suggests our argument in `GSP and GNNs: An overview' can be restated as follows: when trained to predict age, a coVariance filter learns specific ways to exploit the principal components of the anatomical covariance matrix.} The observations here provide a neat link between the computational module in a deep learning model and a PCA-based feature extractor. 
\textcolor{black}{Thus, at least in part, VNNs achieve their learning objective (predicting age here) by exploiting the principal components of the anatomical covariance matrix in a judicious, data-driven manner.}

\begin{myboxii}
    To exhibit reproducibility across independent datasets, it is critical that coVariance filters and VNNs be stable to stochastic perturbations in $\hat\bC_n$ relative to $\bC$.
\end{myboxii}


\textcolor{black}{PCA-driven approaches are prone to unstable or irreproducible inference outcomes as a result of stochastic perturbations in the covariance eigenspectrum due to small changes in the dataset (e.g., by addition or removal of a few samples)~\cite{joliffe1992principal}. VNNs, however, overcome such irreproducibility pitfalls.}

\noindent 
\textcolor{black}{{\bf Stability of VNNs.}} 
The deviation between the outputs of coVariance filters or VNNs for $\hat\bC_n$ relative to $\bC$ is bounded if the frequency response of the coVariance filter is \textcolor{black}{sufficiently smooth in the Lipschitz sense}. \textcolor{black}{This stability result, that we informally state next, implies reproducibility of age prediction outcomes when using a VNN model, unlike comparable PCA-driven approaches (Fig. 6).}

\noindent

\noindent
\begin{theorem}[Stability of coVariance Filters and VNNs (Informal)~\cite{sihag2022covariance}]\label{filterstab}
Consider a random vector $\bx \in \mR^{M}$, such that, $\|\bx\|\leq 1$, and its corresponding ensemble covariance matrix $\bC = \mathbb{E}[(\bx - \mE[\bx])(\bx - \mE[\bx])^{\top}]$. For a sample covariance matrix $\hat\bC_n$ formed using $n$ i.i.d. realizations of $\bx$, if the frequency response satisfies \textcolor{black}{$|h(\phi_i)- h(\phi_j)|\leq \varsigma|\phi_i - \phi_j|$ for an appropriate $\varsigma>0$}, the following holds with high probability:
\begin{align}\label{filterstab_rslt}
    \left\lVert \bH(\hat\bC_n) - \bH(\bC)\right\rVert \leq \alpha_n,
\end{align}
where $\alpha_n$ scales as ${\cal O}(1/n^{1/2-\varepsilon})$ for some $\varepsilon \in(0,1/2)$. Further, for a VNN $\Psi(\cdot;\cdot,{\cal H})$ of depth $L$ and $F$ outputs per MIMO layer, if the pointwise non-linearity $\sigma(\cdot)$ satisfies $|\sigma(a) - \sigma(b)|\leq |a-b|$, then
\begin{align}
    \|\Psi(\bx;\hat\bC_n, \cH)-\Psi(\bx;\bC,\cH)\| \leq LF^{L-1} \alpha_n.
\end{align}
\end{theorem}
\noindent
The eigenvalues $\{\hat\phi_1,\dots, \hat\phi_M\}$ of the sample covariance matrix $\hat\bC_n$ are likely to be perturbed relative to the eigenvalues $\{\phi_1,\dots, \phi_M\}$ of $\bC$. For close eigenvalues of $\bC$, the corresponding estimates in $\hat\bC_n$ may not maintain consistent ordering (in terms of magnitude) with high probability. Hence, a traditional PCA-regression approach is highly vulnerable to irreproducibility when $\hat\bC_n$ is perturbed. However, this concern is mitigated by VNN-based information processing as the filter response $h(\lambda)$ exhibits limited variability for eigenvectors associated with close eigenvalues (see the assumption in Theorem~\ref{filterstab}).





\begin{mdframed}[hidealllines=true,backgroundcolor=gray!20]
\noindent{\bf Improved stability of coVariance filters and VNNs relative to PCA: An age prediction task }

\begin{minipage}{0.99\linewidth}
	\makeatletter
	\def\@captype{figure}
	\makeatother
	\centering
	\vspace{20pt}
	\includegraphics[width=0.9\linewidth]{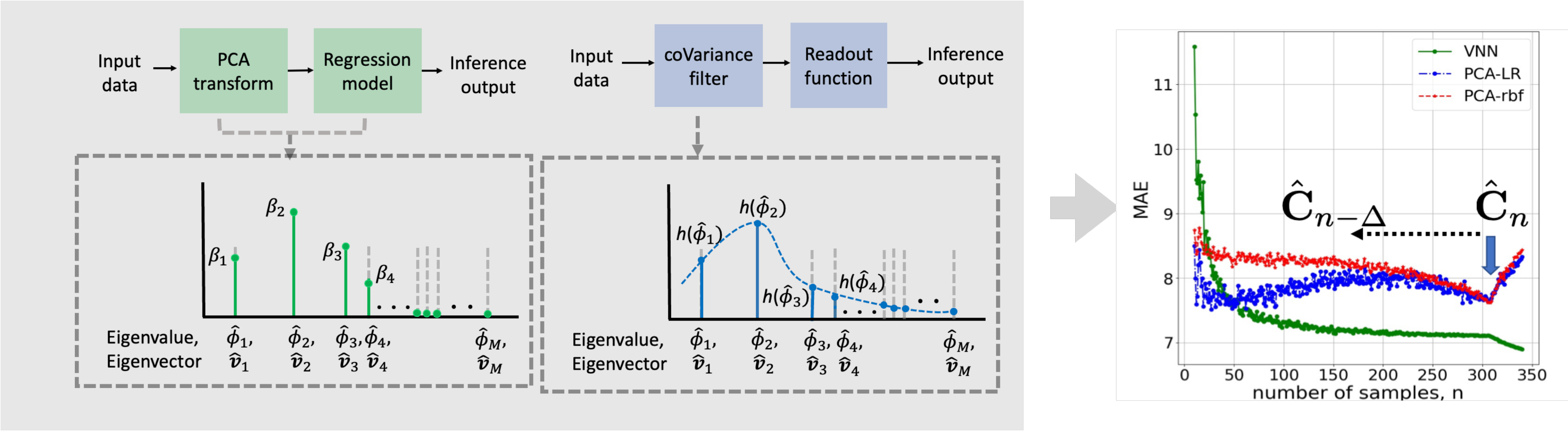}
	\caption{\footnotesize PCA-regression versus a coVariance filter-driven learning approach to regression. PCA-regression explicitly assigns the importance to specific eigenvectors of the sample covariance matrix $\hat\bC_n$. The coVariance filter implicitly determines the influence of specific eigenvectors of $\hat\bC_n$ to the learning task via the \textcolor{black}{frequency} response $h(\hat{\phi}_i)$ for eigenvalues $\hat{\phi}_i$ in $\hat{\boldsymbol\Phi}$. As a consequence of Theorem~\ref{filterstab}, the regression performance of the VNN on the age prediction task using cortical thickness features is consistent even when the sample covariance matrix is perturbed (represented by $\hat\bC_{n-\Delta}$) relative to the one used for training ($\hat\bC_{n}$, marked with a blue arrow)~\cite{sihag2022covariance}. In contrast, PCA-regression with a linear kernel (PCA-LR) or a radial basis function kernel (PCA-rbf) exhibit instability when the principal components are re-evaluated using perturbed sample covariance matrices, while retaining the same regression weights. }		\vspace{20pt}
\end{minipage}

\end{mdframed}

\begin{remark} [Sparsifying the anatomical covariance matrix in VNNs]
\textcolor{black}{Our discussion has so far implicitly assumed that the number of samples $n$ is large enough for $\hat\bC_n$ to be a reasonably accurate approximation of $\bC$. However, \emph{high-dimensional} neuroimaging settings are characterized by small sample sizes, that will adversely affect estimation of the anatomical covariance matrix. This predicament may create memory inefficiency and computational challenges, especially when the true covariance matrix is sparse but $\hat\bC_n$ is dense due to spurious correlations. To address such challenges, \emph{sparse} VNNs have been proposed to attain better quality covariance matrices while preserving the stability of VNNs~\cite{cavallo2024sparsecovarianceneuralnetworks}. Sparse VNNs implement principled hard and soft-thresholding strategies to filter out spurious correlations.}
\end{remark}

\noindent
\textcolor{black}{{\bf Transferability of VNNs.} } VNNs also inherit the GNN property of transference across datasets of different dimensionalities. \textcolor{black}{This makes VNNs compatible with Principle 4 for $\Delta$-Age prediction. The ensuing discussion introduces the mathematical principles behind `successful' transference, i.e., when a VNN retains its performance (without retraining) after being deployed to process another dataset of distinct dimensionality. To further exemplify this desirable property and its practical impact, Fig.~11 illustrates the transferability of VNNs in the context of brain age gap prediction.}

Once more, the theoretical groundwork relies on continuous approximations of discrete objects but now in the VNN setting~\cite{sihagJSTSP}, mimicking the ideas for general GNNs we briefly outlined in `GSP and GNNs: An Overview'. To ground the abstractions, the $M$ partitions of the interval $[0,1]$ to generate $y_{\bx}$ can be interpreted as a partition of the brain cortex into $M$ regions~\cite{sihagJSTSP}. The measures of the $i$-th interval in $y_{\bx}$ and the $i$-th diagonal block in $W_{\bC}$ are chosen to be proportional to the marginal variance $[\bC]_{ii}$; see also~\cite{sihagJSTSP} for technical details. The following theorem establishes transference of a VNN with parameters $\cH$ between datasets of $M_1$ and $M_2$ features. To state the result, let $y_{M_1}$ and $y_{M_2}$ denote the continuous approximations of $\Psi(\bx_{M_1}; \bC_{M_2}, \cH)$ and $\Psi(\bx_{M_2}; \bC_{M_2}, \cH)$, respectively. The notation $\bx_M$ and $\bC_M$ explicitly emphasizes that the number of features is $M$.
\noindent
\begin{theorem}[Transference of VNNs (Informal)~\cite{sihagJSTSP}]\label{transferthm}
Consider two datasets of $M_1$ and $M_2$ features and a VNN $\Psi(\cdot;\cdot,\cH)$ consisting of $L$ layers and $F$ outputs per MIMO layer. If the continuous approximations $W_{\bC_{M_1}}$ and $W_{\bC_{M_2}}$ are close and part of a converging sequence to a suitable graphon limit, and the continuous approximations $y_{\bx_{M_1}}$ and $y_{\bx_{M_2}}$ of the inputs are sufficiently close, then the continuous approximations of the VNN outputs $\Psi(\bx_{M_1};\bC_{M_1},\cH)$ and $\Psi(\bx_{M_2};\bC_{M_2},\cH)$ converge in the sense
\begin{align}\label{transfereq}
\|y_{M_1} - y_{M_2}\|_2 = {\cal O} \left(\frac{1}{M_1^{3\zeta/2- 1} } + \frac{1}{M_2^{3\zeta/2- 1} }\right),\quad \text{for some constant $\zeta\in (2/3,1]$.}
\end{align}
\end{theorem}
Theorem~\ref{transferthm} summarizes the foundational principle behind the transference of VNNs, with tangible impacts to neuroimaging data analysis. \textcolor{black}{In fact, the graphon limit can be viewed as the brain cortex, over which finite-dimensional neuroimaging datasets are sampled. The sampling design is determined by the brain atlas used to divide the cortex into regions of interest. The key upshot of Theorem~\ref{transferthm} is to justify that VNNs are capable of processing datasets curated according to different brain atlases; see also Fig. 11.} 



\begin{mdframed}[hidealllines=true,backgroundcolor=gray!20]

{\bf Transferability of VNNs across multiscale datasets for age prediction}

\begin{minipage}{0.99\linewidth}
	\makeatletter
	\def\@captype{figure}
	\makeatother
	\centering
		\vspace{20pt}
	\includegraphics[width=0.9\linewidth]{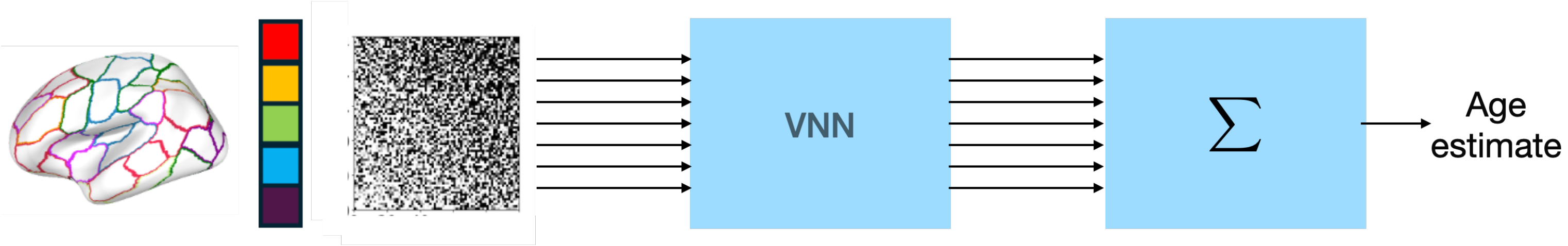}\label{fig_age}
	\caption{\footnotesize VNN model schematic designed for age-prediction with transference across datasets of different dimensionalities (Fig.~5). The readout function is the unweighted mean.
	}	\vspace{20pt}
\end{minipage}

\captionof{table}{Transferability for age prediction task across multiscale datasets (MAE for VNN regression outputs with respect to the ground truth)~\cite{sihagJSTSP}}
{\centering
	{%
		{\footnotesize
			\begin{tabular}{|c| c| c| c|c|}
				\hline
				\multirow{2}{*}{\backslashbox{Training}{Testing}} & \multirow{2}{*}{100 features} & \multirow{2}{*}{300 features} & \multirow{2}{*}{500 features}  \\ 
				& & &  \\
				\hline
				100 features & \cellcolor{red!25}{$5.39 \pm 0.084$} & $5.5 \pm 0.101$ & $5.61 \pm 0.132$ \\
				\hline
				300 features & $5.39 \pm 0.193$ & \cellcolor{red!25}{$5.41 \pm 0.167$} &  {$5.47\pm 0.169$} \\ 
				\hline
				500 features & $5.43 \pm 0.2$ & $5.38 \pm 0.15$ &  \cellcolor{red!25}{$5.4\pm 0.169$} \\
				\hline
		\end{tabular}}
	}\vspace{20pt}
	
The transference of VNN-based regression models across datasets of different dimensionalities is facilitated by setting the readout function to be the unweighted mean (as illustrated in Fig.~7). For the age prediction task on the same healthy population, here we consider multiscale cortical thickness datasets with $M=100$, $300$, and $500$ features spanning the entire brain cortex (curated according to Schaefer's brain atlas)~\cite{sihagJSTSP}. Table 2 reports the regression performance of three VNNs trained on the datasets with $M=100$, $300$, and $500$ features along the diagonal. The performance after transference of the VNNs to a test dataset of different dimensionality is tabulated in the off-diagonal elements. For instance, the element corresponding to the row associated with `$100$ features' and the column associated with `$300$ features' indicates the MAE performance for a VNN trained on the dataset with $100$ features and transferred to the dataset with $300$ features.
}
\end{mdframed}

\section*{Towards Explainable Brain Age Gap Prediction from Structural MRI}
In this section, we close the loop by blending the various technical results and observations from `GSP Foundations and Neuroimaging Data Analysis' to construct a workflow for brain age gap prediction that meets the Principles 1-4. We assume that structural MRI scans are pre-processed with standard pipelines to derive anatomical features~\cite{fischl2012freesurfer}. The key modules of this workflow can be summarized as follows.

\noindent{\bf ML model for $\Delta$-Age prediction.} A VNN is selected as the regression model and \textcolor{black}{the anatomical covariance matrix $\hat\bC_n$ is estimated from the anatomical features $\bx$ of the healthy population. The covariance matrix $\hat\bC_n$ remains fixed in the $\Delta$-Age prediction pipeline.} The \emph{readout} function of the VNN model is chosen to be an unweighted mean function. Hence, the age estimate $\hat y$ is formed by aggregating the learned representations in $\Psi(\bx;\hat\bC_n,\cH)$ (Fig.~7) to yield
\begin{align}\label{vnn_pred}
    \hat y = \frac{1}{M} \sum\limits_{j=1}^M [\bp_{\bf x}]_{j}\;,\quad  \text{where }\quad \bp_{\bf x} = \frac{1}{F} \sum\limits_{f=1}^F [\Psi(\bx;\hat\bC_n,\cH)]_f
\end{align}
and $F$ is the width of the VNN at its final layer. The VNN model is pre-trained to predict the chronological age of a healthy cohort. \textcolor{black}{$\Delta$-Age is derived using the steps described in `How is brain age gap evaluated?'}. The estimates $\hat y$ across the dataset are further corrected for age bias to yield the brain age estimate $\hat y_{\sf B}$ according to~\eqref{s2}, e.g.,~\cite{de2020commentary}, which further provides the estimate of $\Delta$-Age in~\eqref{ss}. 

\textcolor{black}{The information gleaned by the VNN model leading to a higher $\Delta$-Age is embedded in the learned representation $\Psi(\bx; \hat\bC_n, \cH)$. 
Traces of accelerated aging in the anatomical features $\bx$ of an individual with neurodegeneration are encoded as deviations in $\Psi(\bx; \hat\bC_n, \cH)$ relative to what is expected of a healthy individual. These differences propagate downstream through the age estimate $\hat y$ to yield a larger $\Delta$-Age. This transparency on the role of the VNN model for $\Delta$-Age prediction would had been lost if we used a learnable readout function (e.g., a multilayer perceptron). In this hypothesized case, the weights of the VNN and the readout function would \emph{collectively} contribute to the age estimate $\hat y$, leading to conceptual ambiguities. As we proceed in this section, it will become clear how a VNN with an unweighted mean readout function enables a qualitative assessment of $\Delta$-Age, and seamless transference of the brain age gap prediction pipeline across datasets of different dimensionalities. }

\noindent
\textcolor{black}{{\bf Anatomic interpretability of $\Delta$-Age.}}  \textcolor{black}{The anatomic characterization of $\Psi(\bx; \hat\bC_n, \cH)$, how this representation leads to the age estimate $\hat y$ and subsequently, $\Delta$-Age, are key to establishing the desirable \emph{anatomic interpretability} of $\Delta$-Age. To this end, note that the vector $\bp_{\bx}$ is obtained by aggregating across the width of $\Psi(\bx;\hat\bC_n,\cH)$ in~\eqref{vnn_pred}}, \textcolor{black}{such that, the age estimate $\hat y$ is the mean of the elements in $\bp_{\bf x}$. Due to the coVariance filters in the VNN, we can interpret~\eqref{vnn_pred} to state that $\bp_{\bx}$ is formed by transforming the input anatomical features $\bx$ according to the covariance matrix $\hat\bC_n$, where the learnable parameters $\cH$ of the VNN encode the information about healthy aging. When projected on the brain atlas, $\bp_{\bx}$ encodes brain `regional contributions' to the predicted output $\hat y$. For instance, $[\bp_{\bx}]_j$ is the contribution of region $j$.}

Accelerated aging, as captured by $\Delta$-Age, could be hypothesized to be an aggregated effect of anomalous contributions from certain biologically plausible brain regions. \textcolor{black}{Hence, if the representation $\Psi(\bx;\hat\bC_n,\cH)$ encodes the information about accelerated aging, we anticipate that certain elements of $\bp_{\bx}$ will exhibit a `larger contribution' to the age estimate $\hat y$. The \emph{anatomical signatures} of $\Delta$-Age for a neurodegenerative condition can be revealed via group comparisons of appropriately defined statistics of $\bp_{\bx}$, between the cohort with the neurodegenerative condition and a healthy cohort. Direct comparisons between different populations are prone to bias due to differences in the respective chronological age distributions.} To mitigate this bias and better capture accelerated aging in the disease cohort during group-level analyses, we define the `regional residual' statistic for anatomical feature  $j$  (or brain region represented by feature $j$ in this case) with respect to the VNN output $\hat y$ at the regional level as
\begin{align}\label{res_dist}
    [\br]_j \triangleq [\bp_{\bx}]_j - \hat y\;.
\end{align}
%

\noindent
\textcolor{black}{
To understand the contribution of elevated regional residuals to higher $\Delta$-Age for a cohort with accelerated aging, consider a toy example with two individuals of the same chronological age $y$. Suppose that one belongs to the disease group, the other to the healthy cohort. From~\eqref{s2}, their corresponding VNN outputs (denoted by $\hat y_{\sf D}$ for the individual in the disease cohort and $\hat y_{\sf HC}$ for the individual in the healthy cohort) are corrected for age-bias using the same term $\omega y + \varrho$. Hence, $\Delta$-Age for the individual in the disease cohort will be highest only if the VNN prediction $\hat y_{\sf D}$ exceeds $\hat y_{\sf HC}$. Since the VNN predictions  $\hat y_{\sf D}$  and  $\hat y_{\sf HC}$ are proportional to the unweighted aggregations of the regional level estimates [see~\eqref{vnn_pred}], $\hat y_{\sf D}>\hat y_{\sf HC}$ can be a direct consequence of a subset of regional residuals [see~\eqref{res_dist}] being robustly elevated in the disease group relative to the healthy cohort. If the individuals have different chronological ages, the age-bias correction will remove any age-related confounding in the differences in distributions of $\Delta$-Age.}

Based on the arguments above, the brain regions contributing to higher $\Delta$-Age in neurodegeneration can be traced to significantly elevated regional residuals in the disease cohort. Anatomical interpretability thus follows via evaluation of group level differences between the regional residuals of the disease and healthy cohorts (with standard tests such as ANCOVA using variables like age and gender as covariates). The elements of the residual vector $\br$ have been shown to exhibit distinct anatomic signatures for $\Delta$-Age under different neurodegenerative conditions~\cite{sihag2023explainable, sihagisbi}, which could be used for `fingerprinting'.

\noindent {\bf Explaining regional residuals.} Our discussion in `GSP Foundations for Neuroimaging Data Analysis' revealed that learning with VNNs can be understood, at least in part, as a process that implicity exploits the eigenvectors of $\hat\bC_n$. Since the regional residuals are derived directly from the VNN representations, the anomalous behavior in neurodegeneration could be captured (and explained) in terms of how the VNN selectively leveraged said eigenvectors across the different population groups. 
To this end, for an individual with regional residual vector $\br$, we consider the inner product $\bar \br^\top \hat\bv_i$ with eigenvector $\hat\bv_i$ as metric, where $\bar \br $ is the normalized version of $\br$, such that, $\|\bar \br \|_2 = 1$.

\textcolor{black}{Notably, the inner product metric $\bar \br^\top \hat\bv_i$ closely resembles the GFT (here coVariance Fourier transform) in~\eqref{eq9} if the ML model were a coVariance filter. In this case and from~\eqref{eq9}, 
it follows that $\bar \br^\top \hat\bv_i$ 
is a composite metric that combines the frequency response of the coVariance filter $h(\lambda_i)$ and the alignment between the considered eigenvector and the input data, i.e., $\bv_i^{\top} \bx$. The analytical extension of this observation to VNNs is non-trivial. However, since the coVariance filter forms the fundamental computational module in a VNN, we anticipate that the metric $\bar \br^\top \hat\bv_i$ will have different group-level distributions for individuals with neurodegeneration and healthy controls (at least for some eigenvectors). This observation motivates our approach to explainability of $\Delta$-Age in neurodegeneration.} \textcolor{black}{The terms interpretability and explainability are sometimes used interchangeably in the field of explainable AI. Here, we refer to `interpretability' when alluding to anatomical interpretability of $\Delta$-Age. The term `explainablity' is used in the context of understanding how the statistics that yield anatomical interpretability were derived by the VNN. Establishing explainability in this spirit helps disentangle $\Delta$-Age regardless of whether it aligns with the biological hypotheses. Hence, explainability of $\Delta$-Age provides a promising approach not only to support $\Delta$-Age prediction, but also to diagnose unexpected outcomes from a VNN-driven pipeline (for instance, low $\Delta$-Age in a specific set of individuals affected by neurodegeneration).}




\begin{mdframed}[hidealllines=true,backgroundcolor=gray!20]
{\bf Case Study 3: Anatomically interpretable and explainable $\Delta$-Age in AD}

\noindent\textcolor{black}{In this case study, we demonstrate the integration of VNN-driven modules within the $\Delta$-Age prediction workflow for AD. Figure~8 summarizes the workflow for anatomically interpretable $\Delta$-Age prediction using VNNs.}

\begin{minipage}{0.99\linewidth}
	\makeatletter
	\def\@captype{figure}
	\makeatother
	\centering \vspace{20pt}
	\includegraphics[width=0.85\linewidth]{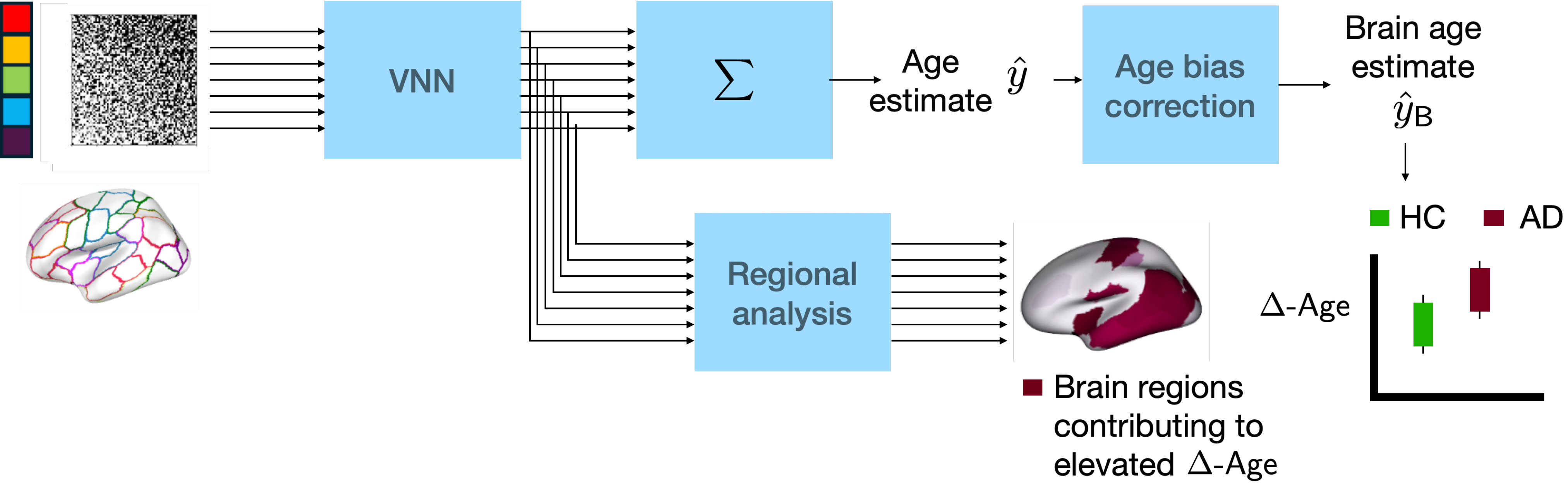}\label{fig_wflow}
	\caption{\textcolor{black}{\footnotesize Workflow for anatomically interpretable $\Delta$-Age prediction using VNNs in the HC and AD cohorts.}}\vspace{20pt}
\end{minipage}

\noindent {\bf VNN model.} Recall the ML model in Case Study 2, namely a VNN trained to predict the chronological age of the HC group from the OASIS-3 dataset~\cite{lamontagne2019oasis}. The model has $L=2$ layers, with the first layer consisting of $2$ filter taps and the second layer consisting of $6$ filter taps. The width is $F=61$. Overall, the VNN model has $22,570$ learnable parameters. The architecture was determined via a hyperparameter optimization procedure using the Optuna package~\cite{akiba2019optuna}. 

\begin{minipage}{0.99\linewidth}
	\makeatletter
	\def\@captype{figure}
	\makeatother
	\centering \vspace{20pt}
	\includegraphics[width=0.65\linewidth]{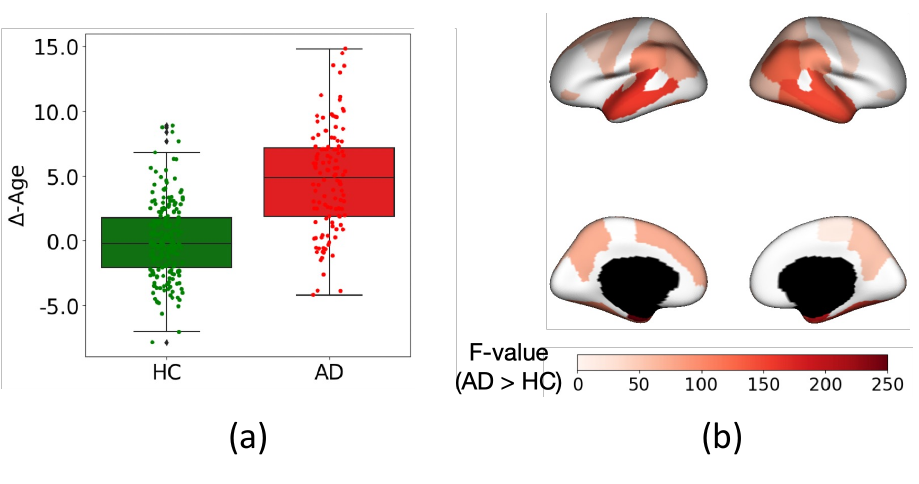}
	\caption{\footnotesize (a) $\Delta$-Age distributions in the HC and AD groups. (b) Higher $\Delta$-Age in AD relative to HC is anatomically interpreted by the group-level analysis of regional residuals for the HC and AD populations.  }\vspace{20pt}
\end{minipage}

\noindent\textcolor{black}{{\bf Anatomically interpretable $\Delta$-Age.} Figure~9a re-illustrates the distributions of $\Delta$-Age for the HC and AD cohorts (see Case Study 2 for additional information). $\Delta$-Age was derived using the same workflow as described in `How is brain age gap evaluated'? Anatomic interpretability via analyses of regional residuals as described in `Towards Explainable Brain Age Gap Prediction from Structural MRI' yielded the anatomical map in Fig.~9b. The $F$-values derived from ANCOVA (age as covariate) from statistically significant group differences ($p$-value after Bonferroni correction for multiple comparisons $<0.05$) in the regional residuals between the HC and AD groups have been projected on the brain surface. Brain regions colored with darker contrast represent the most significant contributors to elevated $\Delta$-Age in the AD group relative to the HC group in Fig.~9a. }

\noindent\textcolor{black}{{\bf Comparing anatomic interpretability with brain atrophy in AD.} Case Study 1 revealed that the AD group exhibited larger brain atrophy relative to the HC group (after controlling for age). Moreover, the atrophy was spread across most brain regions in the AD group, being most statistically significant in the bilateral brain regions spanning the temporal lobe, temporo-parietal junction, and entorhinal regions. Notably, the anatomic interpretability of $\Delta$-Age in Fig. 9b aligned with the brain regions exhibiting the most atrophy in Fig. 1d. Since brain atrophy patterns reflect accelerated aging in structural MRI, we expected an alignment between brain atrophy and anatomic interpretability of $\Delta$-Age. Hence, our experiments attested to the fact that $\Delta$-Age in the AD group was indeed driven by atrophy patterns in structural MRI. This finding is challenging to establish reliably using `black-box' deep learning models. We also note that the anatomic interpretablity supporting elevated $\Delta$-Age in AD (Fig.~9b) was not identical to the brain atrophy patterns in Fig.~1d. Thus, the VNN model refined the information within structural MRI to reveal the key contributors to $\Delta$-Age in AD.}

\begin{minipage}{0.99\linewidth}
	\makeatletter
	\def\@captype{figure}
	\makeatother
	\centering \vspace{20pt}
	\includegraphics[width=0.85\linewidth]{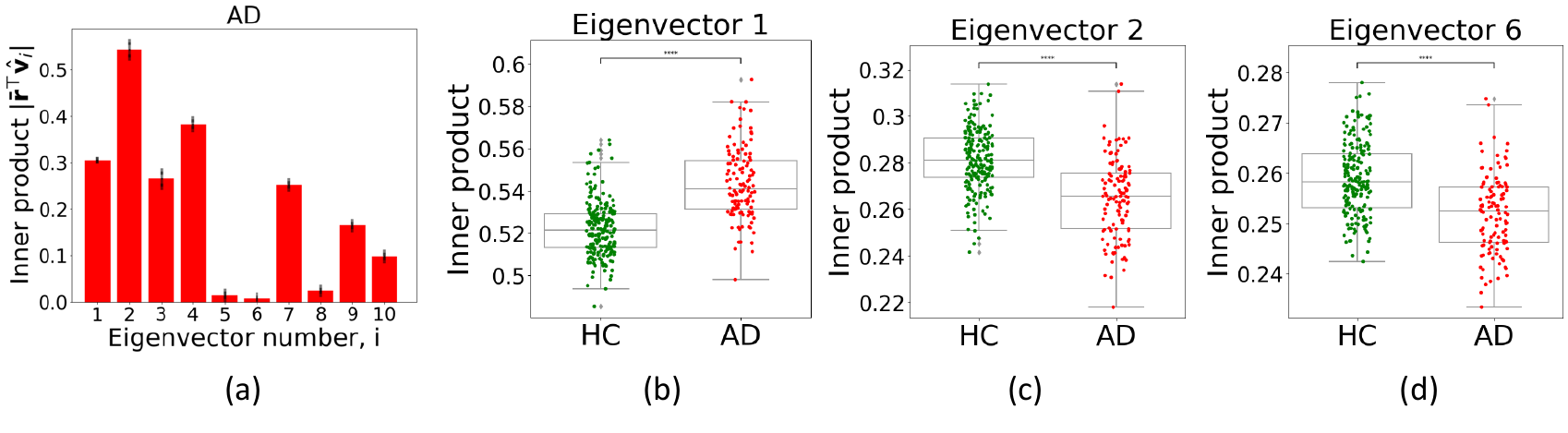}
	\caption{\textcolor{black}{\footnotesize (a) Mean and standard deviation for the inner product metrics $\bar \br^\top \bv_i$ for $i\in \{1,\dots, 10\}$ across the AD group. (b)-(d) The inner product metrics $\bar \br^\top \hat\bv_i$ for eigenvectors 1, 2, and 6 exhibited significant group differences (ANOVA, $p$-value $<0.05$) between the HC and AD cohorts.}}\vspace{20pt}
\end{minipage}

\noindent\textcolor{black}{{\bf Explaining anatomic interpretability of $\Delta$-Age.} 
In Figure~10a, the bars represent the means of the inner product metrics $\bar \br^{\top} \hat\bv_i $ calculated between the first 10 eigenvectors of the anatomical covariance matrix and the normalized regional residuals for the AD group. The whiskers in the bar plot in Fig.~10a illustrate the standard deviations of the inner product metrics across the AD group. The results herein revealed that the eigenvectors of the anatomical covariance matrix exhibited non-uniform importance to $\Delta$-Age, with the first four eigenvectors exhibiting the largest relevance to $\Delta$-Age in AD. Furthermore, the group level comparisons of the inner product metrics between AD and HC groups via ANCOVA (with age as covariate) revealed statistically significant differences observed in the inner product metrics for HC and AD groups for various eigenvectors (results for first, second, and sixth eigenvectors are included in Fig. 10b-d. Altogether, the results in Fig.~10 corroborate the overall relevance of eigenvectors to $\Delta$-Age as well as the relative importance of various eigenvectors for $\Delta$-Age in the AD and HC groups.}

\end{mdframed}

\noindent
Case Study 3 demonstrates how the $\Delta$-Age derived using a VNN model achieves the Principles 1-3 identified earlier. Specifically, by setting the readout function to be an unweighted mean, $\Delta$-Age can be synthesized in terms of VNN-output regional residuals defined at the anatomic level (achieving Principle 1). Elevated $\Delta$-Age in AD vs HC can be traced to regional residual differences at specific brain regions, which are characteristic of AD pathology (such as the medial temporal lobe, among others). The inner product metrics between regional residuals and the eigenvectors of the anatomical covariance matrix revealed how VNNs processed the data for AD,  and how this differed from the HC group (thus, achieving Principle 3). Specifically, during pre-training on the healthy population, the VNN model learned to exploit the eigenvectors of the anatomical covariance matrix in a certain way. This led to specific patterns in regional residuals for the HC population (hence, addressing Principle 2). The regional residuals exhibited distinct behavior for specific brain regions in the AD group. 
     
\begin{mdframed}[hidealllines=true,backgroundcolor=gray!20]
{\bf Case Study 4: Transferability of VNNs validates $\Delta$-Age on multiresolution datasets. }

\begin{minipage}{0.99\linewidth}
	\makeatletter
	\def\@captype{figure}
	\makeatother
	\centering\vspace{20pt}
	\includegraphics[width=\linewidth]{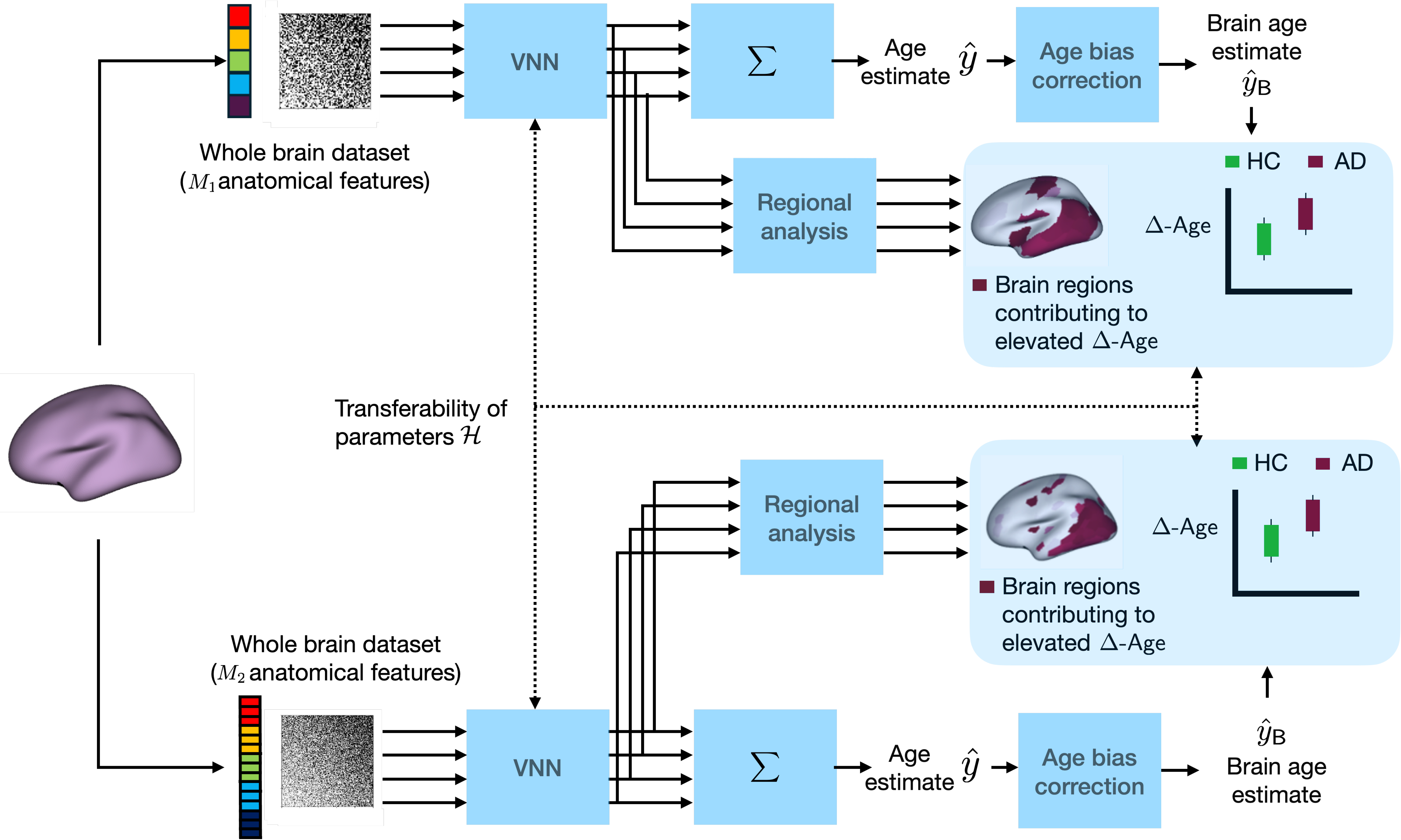}
	\caption{\textcolor{black}{\footnotesize~\cite{sihagJSTSP}. Transferability of VNN yields similar $\Delta$-Age and its associated anatomic interpretability when the VNN model trained on an $M_1$-feature dataset is transferred to study $\Delta$-Age for an $M_2$-feature dataset.}}\vspace{20pt}
\end{minipage}

In this case study, we summarize the results from~\cite{sihagJSTSP}, where the VNN model was shown to exhibit successful transference of $\Delta$-Age and its associated anatomic interpretability across datasets curated to according different versions of the multiscale Schaefer's brain atlas~\cite{schaefer2018local}. The VNN model was trained on cortical thickness features curated according to Schafer's 100 parcellation, 7-network brain atlas. Figure~11 illustrates the transference of VNN model (in terms of its learnable parameters) to yield consistent $\Delta$-Age estimates and associated anatomic interpretability across two cortical thickness datasets of distinct dimensionalities. 

\begin{minipage}{0.99\linewidth}
	\makeatletter
	\def\@captype{figure}
	\makeatother
	\centering\vspace{20pt}
	\includegraphics[width=\linewidth]{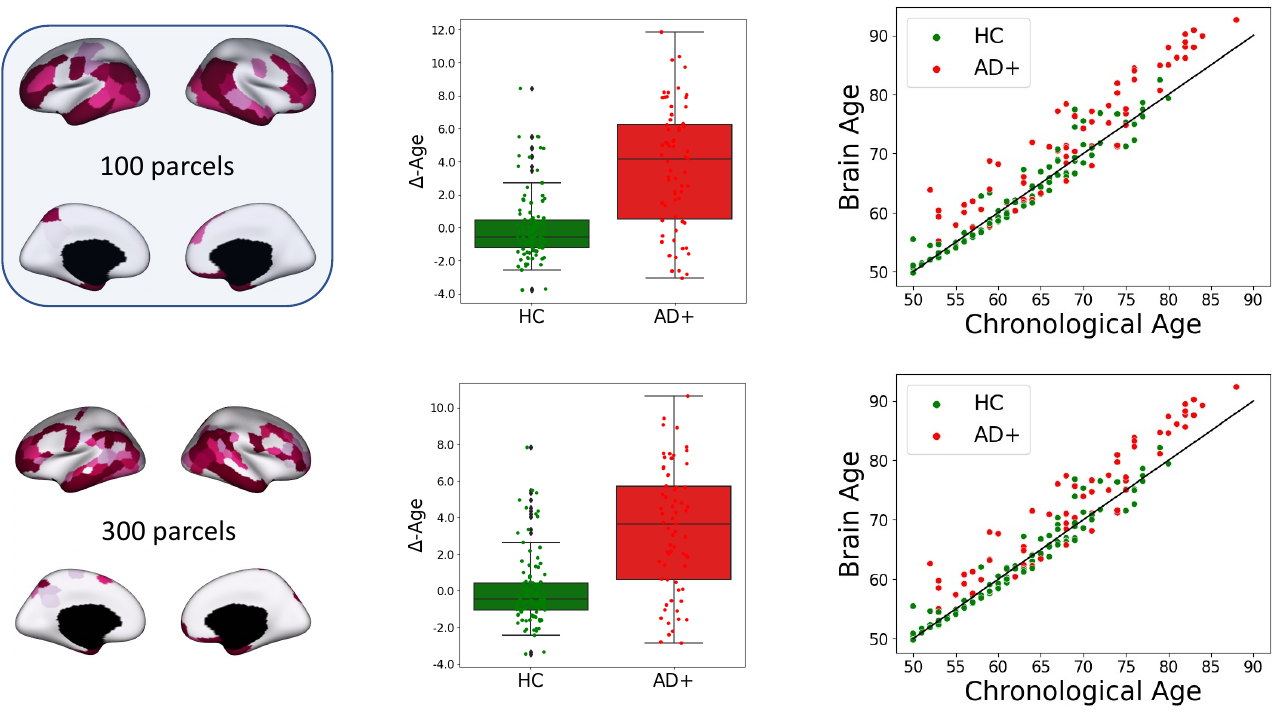}
	\caption{\footnotesize~\cite{sihagJSTSP}. Empirical validation of transference of VNNs across cortical thickness datasets curated according to Schaefer's brain atlas for $\Delta$-Age prediction and associated anatomical interpretability.}\vspace{20pt}
\end{minipage}

Successful transference of VNNs in this context is supported by the theoretical properties of VNNs (Theorem~\ref{transferthm} and associated conditions). The anatomical covariance matrices for the datasets curated according to $100$-features or $300$-features brain atlases can be shown to be part of a converging sequence~\cite{sihagJSTSP}, corroborating the findings in Fig. 12.	
\end{mdframed}

Case Study 4 offers empirical support to the theoretical result in Theorem~\ref{transferthm}.  Under certain regularity conditions, the $\Delta$-Age and its associated anatomic interpretability inferred by a specific VNN model can be transferred across neuroimaging datasets of different dimensionalities (thus, achieving Principle 4). 

\noindent
\textcolor{black}{{\bf Explainable $\Delta$-Age prediction beyond AD.} We have demonstrated various facets of a VNN-driven $\Delta$-Age prediction pipeline by focusing on AD. However, the ML model is broadly applicable as it does not depend on a specific neurodegenerative disease. In fact, the findings in~\cite{sihagisbi} indicate that VNNs can generate an anatomically interpretable $\Delta$-Age in FTD, CBS, and PSP conditions. These neurodegenerative conditions exhibit brain atrophy patterns different from AD and therefore, their anatomical signatures differed from Fig.~9b. For instance, the VNN-driven $\Delta$-Age prediction in FTD revealed both frontal and temporal regions as contributors to FTD in~\cite{sihagisbi}, unlike for AD in Fig.~9b, where temporal regions were prominently implicated. }

\textcolor{black}{Furthermore, the VNN-driven $\Delta$-Age pipeline can also yield clinical insights into cognitively healthy individuals. Specifically, in~\cite{sihag2024brain}, $\Delta$-Age was reported to be correlated with plasma neurofilament light chain (NfL) for a cohort of amyloid-positive and cognitively healthy individuals in the ADNI dataset. Amyloid positivity has been linked with accelerated cognitive decline~\cite{janssen2022characteristics} and plasma NfL is a promising blood biomarker of axonal degeneration~\cite{mielke2019plasma}. Notably, the correlation between $\Delta$-Age and plasma NfL in this cohort was driven by the regional residuals in bilateral entorhinal regions, which are implicated in the early stages of AD pathology~\cite{braak1991neuropathological}. Thus, the discussion here provides promising evidence to support the broader impacts of VNN-driven $\Delta$-Age towards understanding neurodegenerative conditions in various stages of the disease.} 

\section*{Conclusions and Future Outlook}
Brain age gap is a promising ML-driven biomarker derived from neuroimaging data, which has not yet been widely adopted in practice due to several methodological obscurities. In this tutorial, we highlighted the major challenges facing prevalent approaches and concluded that performance-driven methods are inadequate for the practical viability of this application. Our focus has been primarily on structural MRI, because it is the most widely adopted neuroimaging modality in clinical applications.  Brain age prediction algorithms designed for other neuroimaging modalities exhibit similar shortcomings. In this context, we identified four key mathematical principles that could embellish the practical viability of brain age gap prediction. Broadly, we argued for a shift in focus towards brain age gap instead of brain age, for qualitative (and not performance-driven) assessments of regression models trained on a healthy population, and for the generalizability of $\Delta$-Age to different collections of anatomical features derived from structural MRI.

Hence, an amalgamation of mathematical principles and operational requirements of neuroimaging data analysis is critically needed to address the current limitations facing $\Delta$-Age prediction. To this end, we identified GSP as the key analytical tool to enable principled prediction of $\Delta$-Age. GSP-driven learning architectures benefit from improved interpretability of learning outcomes in terms of spectral representations of the graph structure, as well as much-needed theoretical guarantees on robustness and generalizability. 
We surveyed a mix of theoretical results on VNNs and case studies to highlight the steps for the principled construction of $\Delta$-Age, its anatomic interpretability, explainability, and generalizability. \textcolor{black}{Admittedly, the robustness of age prediction models to factors such as distribution shifts is key to achieve reproducible outcomes for different neurodegenerative cohorts. A holistic qualitative performance analysis along with a theoretical understanding of the ML model responsible for $\Delta$-Age prediction is much needed.} 

Looking ahead, brain age gap prediction is a promising tool with a potentially transformative translational impact on digital health and precision medicine. One of the most attractive characteristics of this general approach is its wide applicability to a variety of neurodegenerative conditions. We contend that the underlying ML models possess a characteristic similar to a foundation model, where they can transfer the information learned from healthy aging to yield meaningful biomarkers for various neurodegenerative conditions. Hence, brain age gap prediction algorithms can inform the development of domain-specific foundation models for brain health assessments in the clinic. Our perspective in this article is to also present GSP as a valuable analytical tool in this relatively unexplored context. \textcolor{black}{Extending the principles surveyed in this paper to richer anatomical networks (such as morphometric similarity networks~\cite{seidlitz2018morphometric}) for brain age gaps prediction is a promising future direction.}

Characterizing heterogeneity within disease populations has become increasingly relevant recently, as it can enable targeted interventions and therapies. 
We argued that GSP-informed architectures facilitate seamless integration of information within anatomical features derived from structural MRI, aging, and the anatomical covariance matrix, to yield representations predictive of accelerated aging in neurodegeneration. More broadly, we believe that GSP principles could be fruitfully leveraged to unveil heterogeneous impacts of neurodegeneration via subtyping of clinically-relevant populations. 

In summary, GSP tools hold great promise in becoming one of the dominant analytic paradigms driving ongoing pursuits in digital health and precision medicine, with tangible impacts in the near future. The theoretical foundation of GSP-driven ML models lends unparalleled depth and reliability to their outcomes. This article provides a compelling example of how ML theory can inform major conceptual advancements in the design of data-scientific and application-relevant neuroimaging solutions. 



\section*{Acknowledgement}
Data used in Case Studies 1 and 2 were obtained from the Alzheimer's Disease Neuroimaging Initiative (ADNI) database (adni.loni.usc.edu).  As such, the investigators within the ADNI contributed to the design and implementation of ADNI and/or provided data but did not participate in the analysis or writing of this article. Data collection and sharing for the Alzheimer's Disease Neuroimaging Initiative (ADNI) is funded by the National Institute on Aging (National Institutes of Health Grant U19AG024904). The grantee organization is the Northern California Institute for Research and Education. In the past, ADNI has also received funding from the National Institute of Biomedical Imaging and Bioengineering, the Canadian Institutes of Health Research, and private sector contributions through the Foundation for the National Institutes of Health (FNIH). 

The brain plots were created using the `fsbrain' package in R~\cite{schafer2020fsbrain}.
\section*{Biographies}
\noindent\textbf{Saurabh Sihag} is an Assistant Professor in the Department of Electrical and Computer Engineering at the University at Albany. He received his PhD degree in Electrical Engineering from Rensselaer Polytechnic Institute, in 2020. He has previously been the recipient of J. Baliga fellowship and Charles M. Close ’62 Doctoral Prize for his doctoral dissertation. His research interests include statistical signal processing, network neuroscience, machine learning, and information theory.  

\noindent{\bf Gonzalo Mateos} received his B.Sc. degree in Electrical Engineering from Universidad de la Republica, Montevideo, Uruguay in 2005 and the M.Sc. and Ph.D. degrees in Electrical Engineering from the University of Minnesota, Minneapolis, in 2009 and 2012. Currently, he is a Professor with the Department of Electrical and Computer Engineering, University of Rochester, as well as the Associate Director for Research at the Goergen Institute for Data Science and Artificial Intelligence. He also was an Asaro Biggar Family Fellow in Data Science (2020-23). His research interests lie in the areas of statistical learning from complex data, network science, decentralized optimization, and graph signal processing.

\noindent{\bf Alejandro Ribeiro} received the B.Sc. degree in Electrical Engineering from the Universidad de la Rep\'ublica Oriental del Uruguay, Montevideo, Uruguay, in 1998, the M.Sc. and Ph.D. degrees in Electrical Engineering from the Department of Electrical and Computer Engineering, University of Minnesota, Minneapolis, MN, USA, in 2005 and 2007, respectively. Since 2008, he has been with the University of Pennsylvania (Penn), Philadelphia, PA, USA, where he is currently a Professor of Electrical and Systems Engineering. His research interests include the applications of statistical signal processing to the study of networks and networked phenomena, structured representations of networked data structures, graph signal processing, network optimization, robot teams, and networked control. 

{\tiny 
\bibliographystyle{ieeetr}
\bibliography{spm_bib.bib}
}
\vfill

\end{document}